\documentclass[useAMS,usenatbib]{mn2e}
\newcommand{\mum}{\,$\umu$m}
\usepackage{times}
\usepackage{graphicx}
\usepackage{booktabs}
\usepackage{upgreek}
\usepackage{soul}
\usepackage{color}

\title{SCUBA-2: on-sky calibration using submillimetre standard sources}
\author[J. T. Dempsey et al.]
{\parbox{\textwidth}{J. T. Dempsey$^{1}$\thanks{E-mail:j.dempsey@jach.hawaii.edu}, 
P. Friberg$^{1}$,
T. Jenness$^{1}$, 
R.~P. J. Tilanus$^{1}$,
H.~S. Thomas$^{1}$, 
W.~S. Holland$^{2,3}$, 
D. Bintley$^{1}$,
D.~S. Berry$^{1}$, 
E.~L. Chapin$^{1,4}$\thanks{Present address: XMM SOC, ESAC, Apartado 79, 28691 Vil-
aneueva de la Canada, Madrid, Spain}, 
A. Chrysostomou$^{1}$, 
G.R. Davis$^{1}$, 
A. G. Gibb$^{4}$, 
H. Parsons$^{1}$, 
E. I. Robson$^{2}$}\vspace{0.4cm}\\
\parbox{\textwidth}{$^{1}$Joint Astronomy Centre, 660 N. A'ohoku Place, University Park, Hilo, Hawaii 96720, USA\\
$^{2}$UK Astronomy Technology Centre, Royal Observatory, Blackford Hill, Edinburgh, EH9 3HJ\\
$^{3}$Institute for Astronomy, University of Edinburgh, Royal Observatory, Blackford Hill, Edinburgh, EH9 3HJ\\
$^{4}$Department of Physics and Astronomy, University of British Columbia, 6224 Agricultural Road, Vancouver BC V6T 1Z1, Canada}}

\begin{document}

\date{}

\pagerange{\pageref{firstpage}--\pageref{lastpage}} \pubyear{2012}

\maketitle

\label{firstpage}

\begin{abstract}
SCUBA-2 is a 10000-bolometer submillimetre camera on the James Clerk Maxwell Telescope (JCMT). The instrument commissioning was completed in September 2011, and full science operations began in October 2011. To harness the full potential of this powerful new astronomical tool, the instrument calibration must be accurate and well understood. To this end, the algorithms for calculating the line-of-sight opacity have been improved, and the derived atmospheric extinction relationships at both wavebands of the SCUBA-2 instrument are presented. The results from over 500 primary and secondary calibrator observations have allowed accurate determination of the flux conversion factors (FCF) for the 850 and 450\mum\ arrays. Descriptions of the instrument beam-shape and photometry methods are presented. The calibration factors are well determined, with relative calibration accuracy better than 5 per cent at 850\mum\ and 10 per cent at 450\mum, reflecting the success of the derived opacity relations as well as the stability of the performance of the instrument over several months. The sample-size of the calibration observations and accurate FCFs have allowed the determination of the 850 and 450\mum\ fluxes of several well-known submillimetre sources, and these results are compared with previous measurements from SCUBA.
\end{abstract}


\section{Introduction}

Ground-based submillimetre astronomy has progressed markedly in the last fifteen years. SCUBA-2 \citep{holland2012} is the successor to the highly successful SCUBA instrument \citep{holland1999} which was operational on the JCMT from 1998 to 2005. In both cases, a limitation of ground-based submillimetre observing lies in the attenuation and distortion of the incoming radiation by the atmosphere. The dominant source of this attenuation is absorption by water vapour which, even on high sites such as Mauna Kea, limits submillimetre observations to a few narrow wavebands between the strongest absorption lines.\\

The opacity in these bands can vary rapidly and markedly as a function of precipitable water vapour (PWV), particularly at shorter wavelengths. Rapid measurements of the line-of-sight PWV are required in order to compensate adequately for this attenuation in the observation. A model of the atmospheric transmission as a function of wavelength must be used to relate the PWV to the opacity in the filter bands of the camera. This relation must be determined accurately. Uncertainties affect the short-wavelength windows significantly as they are more opaque and are more highly attenuated as the weather becomes poorer. These atmospheric effects have been investigated previously for SCUBA \citep{archibald} and for Bolocam \citep{sayers}. \\

The second parameter required to calibrate submillimetre continuum data is a measure of the optical throughput of both the telescope and instrument itself. This value is referred to as the flux conversion factor (FCF) which converts the measured power in picowatts (pW) into astronomical fluxes in janskys (Jy). To obtain this factor, a series of astronomical sources of known flux are observed, preferably in a range of atmospheric conditions and over an extended time. Variations in this factor are dominated by optical effects from thermal distortions of the dish (particularly in the early evenings). Poor focus or even submillimetre seeing in the atmosphere will also result in deviations in the peak flux of a compact source.\\

This paper describes the atmospheric transmission relations for both SCUBA-2 filter bands and how these are applied to the data. The calibration sources are listed and the methods of observation and data reduction are described. The results of aperture photometry used to calculate the flux conversion factors are presented. The beam shapes at both wavelengths are modeled and discussed. Finally, a number of well-known submillimetre sources are presented with their calibrated fluxes at 850\mum\ and 450\mum. These values are compared with previous results measured with SCUBA.\\

\section[]{SCUBA-2}
The commissioning of SCUBA-2 was completed on the JCMT in September 2011. The camera observes in two wavebands, 850 and 450\mum\ simultaneously, with 5120 pixels in each focal plane. The full description of the instrument and its performance characteristics are described in \cite{holland2012}. Results from the instrument characterisation and on-sky commissioning are presented in \cite{bintley2010,bintley2012} and \cite{dempsey2012}.  The SCUBA-2 bolometers are integrated arrays of superconducting transition-edge sensors (TESs).  The devices are voltage biased to a temperature near the centre of their transition range, whereupon any temperature fluctuation results in a current change through the TES. The resulting magnetic field change is then amplified using chains of superconducting quantum interference devices (SQUIDs) and the amplified current is digitised. SCUBA-2 detectors have individually coupled resistive heaters which are used to compensate for changes in incident optical power, thereby allowing the detector bias point to remain constant over a wide-range of sky powers. 

\subsection{Flat-fielding}
\begin{figure}
   \begin{center}
   \begin{tabular}{c}
   \includegraphics[height=4cm]{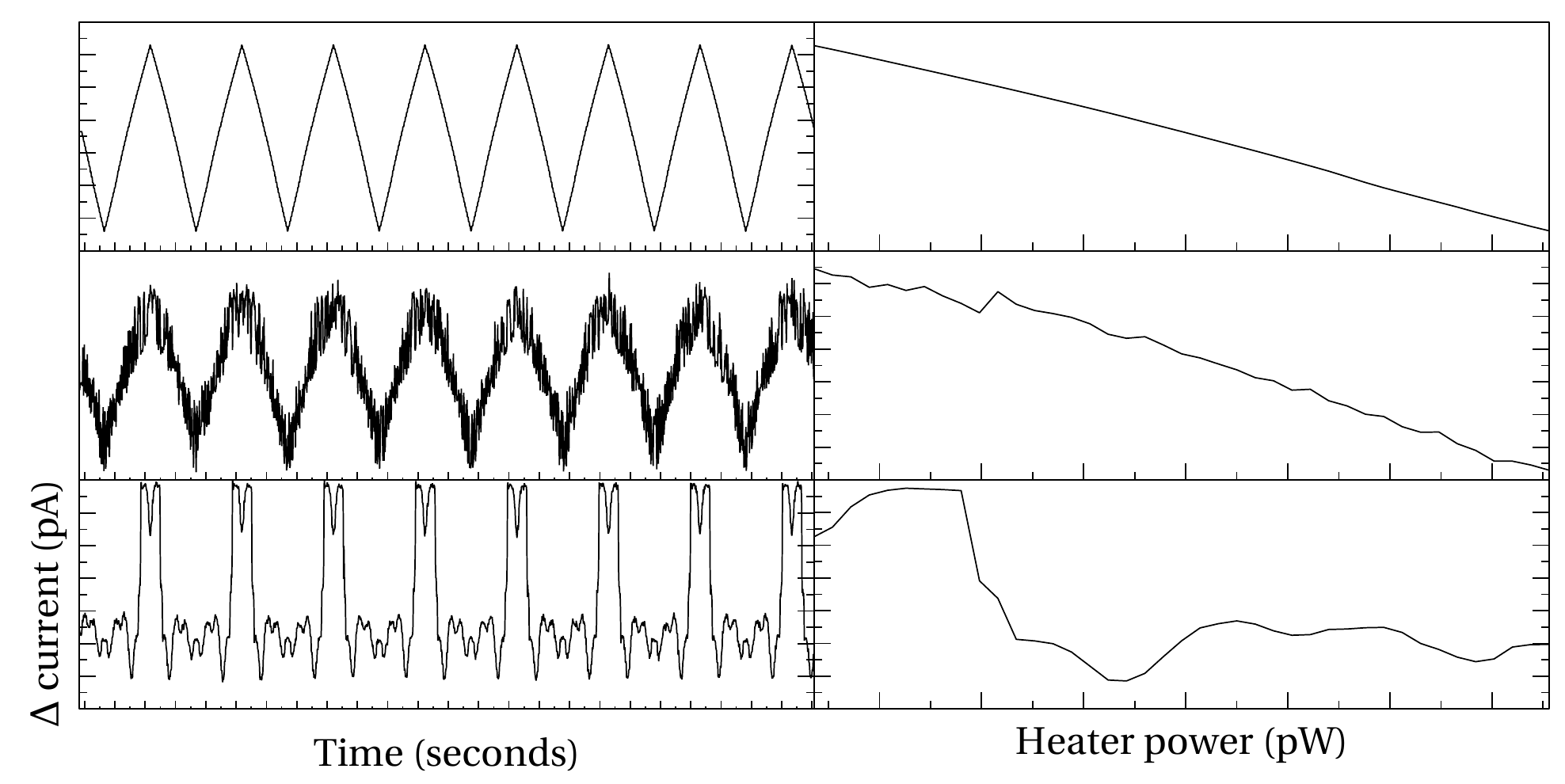}
   \end{tabular}
   \end{center}
   \caption{Examples of the $\sim$3 pW peak-to-peak fast ramps of the heater power that are used to calculate the flat-field response (right) of each individual TES bolometer. The heater is ramped eight times in a ten second interval. The top plots show a typical, well behaved bolometer ramp and the corresponding current change as a function of heater power, with the responsivity determined from a fit to the inverted slope $dI/dP$. Bolometers are flagged and removed as bad if showing too poor a S/N (middle plots) or a severe non-linear response (bottom). \label{fig:flatfield} 
}
   \end{figure} 
 
On-sky observations with SCUBA-2 proceed following an array setup, which is conducted with the internal instrument shutter closed to optimise the SQUID settings.  When the shutter is opened the power from the resistive heater coupled to each bolometer is decreased to compensate for the incident optical power. This adjustment is termed a ``heater track''. The resistor coupling to the TES varies between sub-arrays. The ``effective'' resistance can be calculated by measurement of the I-V curve obtained at different heater power settings. These resistances were shown to be significantly different for each individual sub-array. The relative optical response of the arrays was determined by measurement of the signal when observing bright sources such as Jupiter. The coupling factors were adjusted such that the picowatt response to the planet signal was equal in all sub-arrays. The I-V curves were used to obtain the average absolute pW calibration of the bolometers and then the flux-determined factors were used to calibrate the relative sub-array response, as this is ultimately the measure of the optical response through the entire system from an astronomical source.\\

The heaters are then employed to calibrate the individual bolometer responsivity. The response of the detectors is determined by measuring the slope of the current/power change as the heater power below each pixel is ramped quickly over a small range (a few pW peak-to-peak), as shown in the left-hand plots in Figure~\ref{fig:flatfield}. The flat-field response is the inverse gradient of the current change as a function of heater power (or $dI/dP$) of each individual bolometer. At this point the quality of the bolometer performance is assessed: bolometers with responsivities above or below a threshold limit are removed (these limits are extreme, and are applied to remove the handful of bolometers which display responsivities too high or low to be physically accurate), along with bolometers whose response, $dI/dP$, is non-linear, or shows too poor a S/N. Figure~\ref{fig:flatfield} shows a good bolometer trace (top) and example of a bolometer that was flagged resulting from a poor S/N (centre) and extreme non-linear response (bottom). A flat-field is performed at the start of every science observation, and the quality flags typically remove less than 10 per cent of the total bolometer yield at 450\mum\ and less than 20 per cent of the array yield at 850\mum\ with the percentage and pattern of the subsequent masking being typically consistent.\\

Initially, this flat-fielding was completed in the dark, with the internal instrument shutter closed, assuming that the heater track as the shutter opened would compensate sufficiently for the change in optical power. Testing showed, however, that better bolometer yields and more stable operation was achieved if this flat-fielding was conducted again once the shutter was open, particularly when there was sky instability. Dark flat-fields are still conducted at the end of an array setup to check if the setup was successful.\\ 

\section{Submillimetre atmospheric transmission}\label{trans}

\begin{figure}
   \begin{center}
   \begin{tabular}{c}
   \includegraphics[height=8cm]{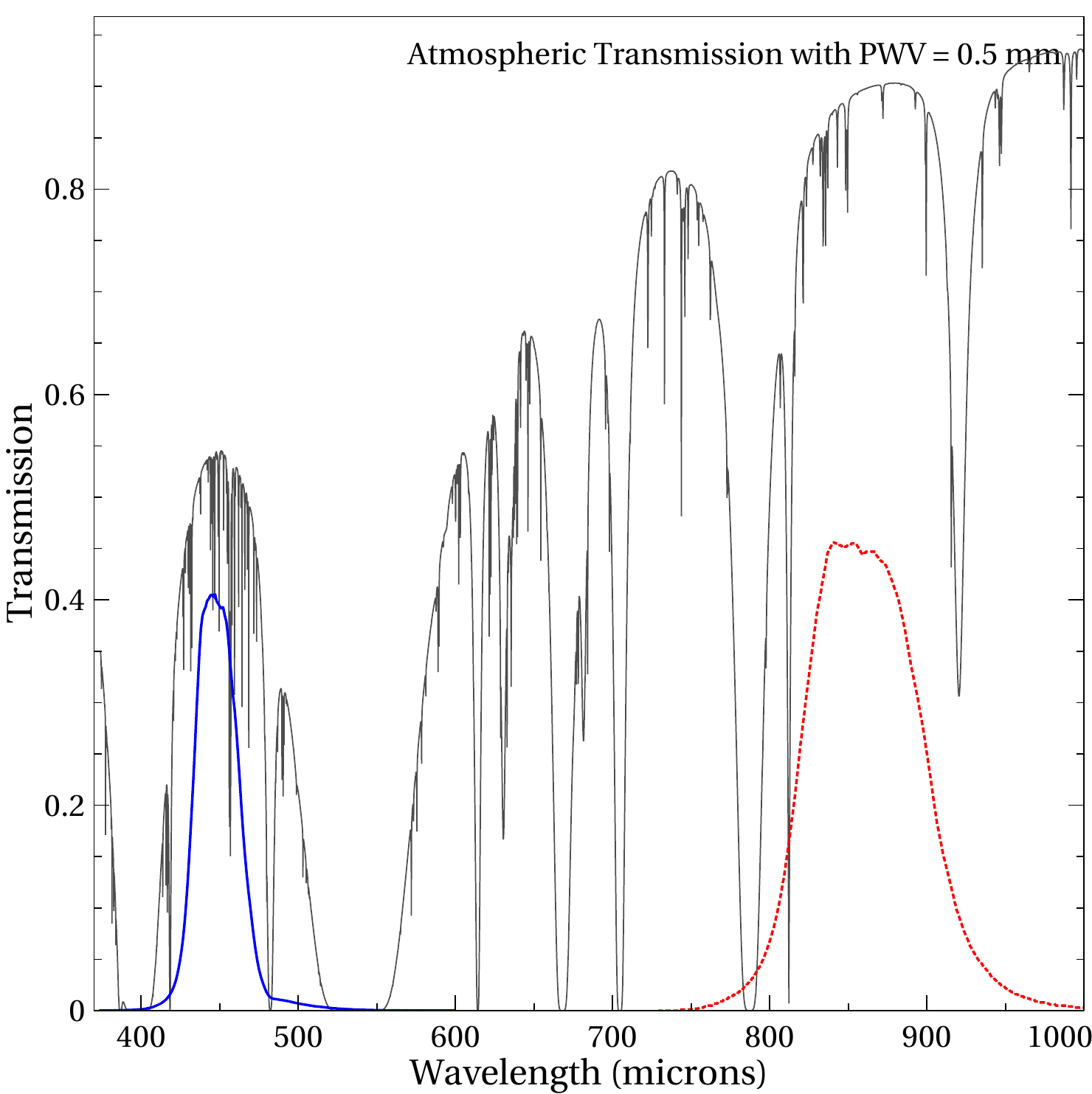}
   \end{tabular}
   \end{center}
   \caption[example]
   { \label{fig:atm}
450\mum\  (blue solid) and 850\mum\ (red dotted) SCUBA-2 filter profiles superimposed on the submillimetre atmospheric transmission curve for Mauna Kea assuming 0.5$\,$mm of precipitable water vapour.}
   \end{figure}

The atmosphere limits ground based observations at submillimetre wavelengths, providing only a few semi-transparent windows even at a high, dry site such as the summit of Mauna Kea. SCUBA-2, like its predecessor SCUBA, has been designed to take advantage of the 850 and 450\mum\ atmospheric windows. The measured SCUBA-2 filter profile at both wavelengths are shown superimposed on the submillimetre atmospheric transmission windows in Figure~\ref{fig:atm}. The filter profiles were measured using a Fourier transform spectrometer on a single bolometer at a time. The filter profiles were measured on a number of bolometers moving away from the centre of the focal plane and no discernible difference was observed in the profile.\\

Atmospheric attenuation at a given waveband in the submillimetre can be described as follows, assuming a plane-parallel atmosphere:
\begin{equation}
I_{\rmn m} = I_{\rmn 0} \;\rmn{e}^{-{\uptau} A},
\label{eq:atm}
\end{equation}
where $I_{\rmn m}$ and $I_{\rmn 0}$ are the measured signal at the telescope and at the top of the atmosphere respectively, $\uptau$ is the extinction coefficient, and $A$ is the airmass (eg. \cite{stevens}). To accurately reconstruct the un-attenuated signal $I_{0}$, the extinction coefficient $\uptau$ must be well described. At the shorter of SCUBA-2's filter bands a 20 percent error in $\uptau$ can result in a corresponding flux  error of 50-80 percent \citep{archibald}.\\

\subsection{The Water Vapour Monitor}

The extinction coefficient, $\uptau$,  must be explicitly determined for each observing waveband. The Caltech Submillimetre Observatory (CSO) has a fixed-azimuth 225$\,$GHz tipping radiometer that has been in operation throughout the lifetime of both SCUBA and SCUBA-2 \citep{radford}. This completes a skydip every fifteen minutes and calculates a zenith opacity using a plane-parallel atmospheric model. The wavelength dependent extinction relations for SCUBA were calculated as a function of the CSO tipper 225$\,$GHz opacity \citep{archibald}. The JCMT has a 183$\,$GHz water vapour monitor (WVM) installed in the receiver cabin that measures the precipitable water vapour (PWV) along the line-of-sight at 1.2-second intervals \citep{wiedner}. The WVM has the advantage of better time resolution, which is important given that SCUBA-2 collects data at a rate of 200$\,$Hz. Also, by measuring directly along the line-of-sight of the telescope we are not limited to the fixed azimuth of the CSO tipper, and we do not need to assume a plane-parallel atmosphere. Skydips taken with the 345$\,$GHz HARP instrument \citep{buckle} on the JCMT, as well as previous comparisons between SCUBA skydips and the CSO tipper show that the atmosphere deviates from a plane-parallel approximation about 30 per cent of the time, particularly in the early evening when the atmosphere is unstable \citep{dempsey2008}. \\

\begin{figure}
  \begin{center}
    \begin{tabular}{c}
      \includegraphics[height=4.2cm]{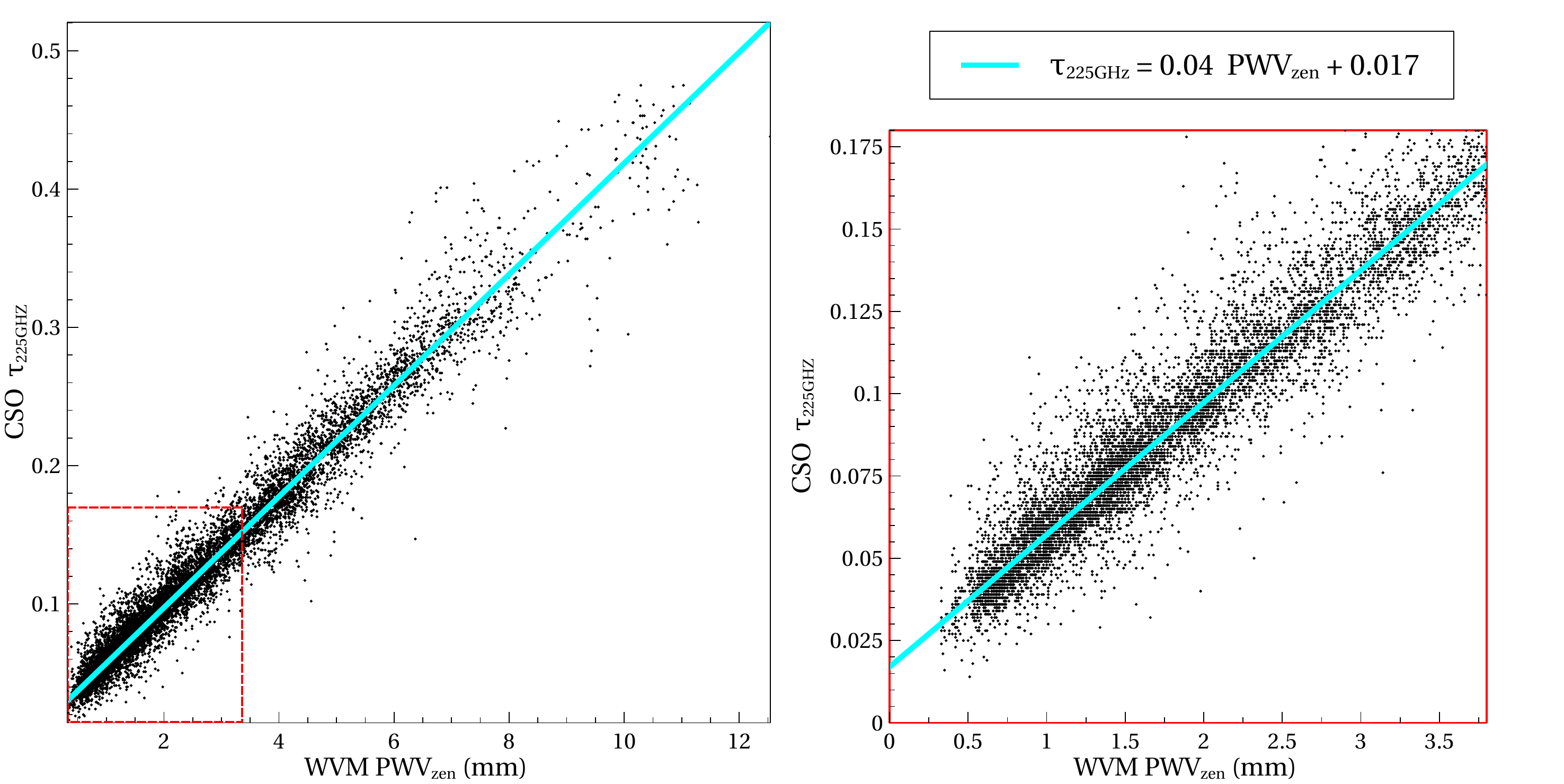}
   \end{tabular}
  \end{center}
  \caption[atm]
  { \label{fig:cso}
     CSO 225$\,$GHz radiometer data vs precipitable water vapour (in millimetres) as derived by the 183$\,$GHz line-of-sight water vapour monitor at JCMT. WVM and CSO data collected in the most stable part of the night, between 9$\,$pm and 3$\,$am, is shown. The full range of PWV fitted is shown on the left, the red rectangle highlighting the best weather conditions ($<$ 3$\,$mm PWV) which is shown in detail in the right hand plot.}

\end{figure}

If we adjust the PWV to its zenith value, PWV$_{\mbox{\tiny zen}}$ by dividing by the airmass, there exists a simple linear relation between it and the opacity at 225$\,$GHz. We can then directly compare the opacity derived in this way by the WVM to the independently produced CSO 225$\,$GHz opacity. The WVM and CSO values show excellent correlation for the stable part of the night between 9$\,$pm and 3$\,$am, with deviations occurring in the early evening and at sunrise when conditions become more variable. The PWV$_{\rmn{zen}}$ is related to  the opacity at 225$\,$GHz, $\uptau{\rmn{\scriptsize (225)}}_{\rmn{wvm}}$ by : 
\begin{equation}
\uptau{\rmn{\scriptsize (225)}}_{\rmn{ wvm}} = 0.04 \; {\rmn{PWV}}_{\rmn{zen}} + 0.017.
\end{equation}

where PWV$_{\mbox{\tiny zen}}$ is in units of millimetres. The comparison between the CSO 225$\,$GHz opacity at zenith and the PWV$_{\rmn{zen}}$ from the WVM for a full year of observing in 2011 is shown in Figure~\ref{fig:cso}.\\

\subsection{Tau relations}

The extinction coefficients at each SCUBA-2 wavelength will be described in terms of the PWV determined by the water vapour monitor. The relation was derived by collating the calibration observations of sources of known flux taken over the six month period from 2011 May until 2011 October, and then in additional science time until 2012 May. These observations were reduced without applying any assumed extinction correction and then fitted using a least-squares algorithm to the following formula, where the precipitable water vapour (PWV) is measured in millimetres:
\begin{equation}
\uptau_{\lambda} =  a \; ({\mbox{PWV}} - b).
\end{equation} 

The resulting extinction coefficients at each wavelength are:
\begin{equation}
\uptau_{850} =  0.179 \;({\mbox{PWV}} + 0.337)
\end{equation} 
and
\begin{equation}
\uptau_{450} =  1.014 \;({\mbox{PWV}} + 0.142).
\end{equation} 

The data from which these extinction coefficients were determined is presented and discussed in Section~\ref{results} below. The quality of these fits is thus determined by the scatter and systematics in the calibrator data, and is accounted for in the resulting error quoted for the flux conversion factor in Section~\ref{results}. The SCUBA-2 extinction correction coefficients in terms of the opacity at 225$\,$GHz are: $\uptau_{\mbox{\tiny 850}} =  4.6 \;(\uptau_{\mbox{\tiny 225}} - 0.0043)$ and $\uptau_{\mbox{\tiny 450}} =  26.0 \; (\uptau_{\mbox{\tiny 225}} - 0.012)$, to allow easy comparison with the previous extinction terms for SCUBA.

\section{SCUBA-2 beam shape}\label{beam}

\begin{figure}
  \begin{center}
   \includegraphics[height=8cm]{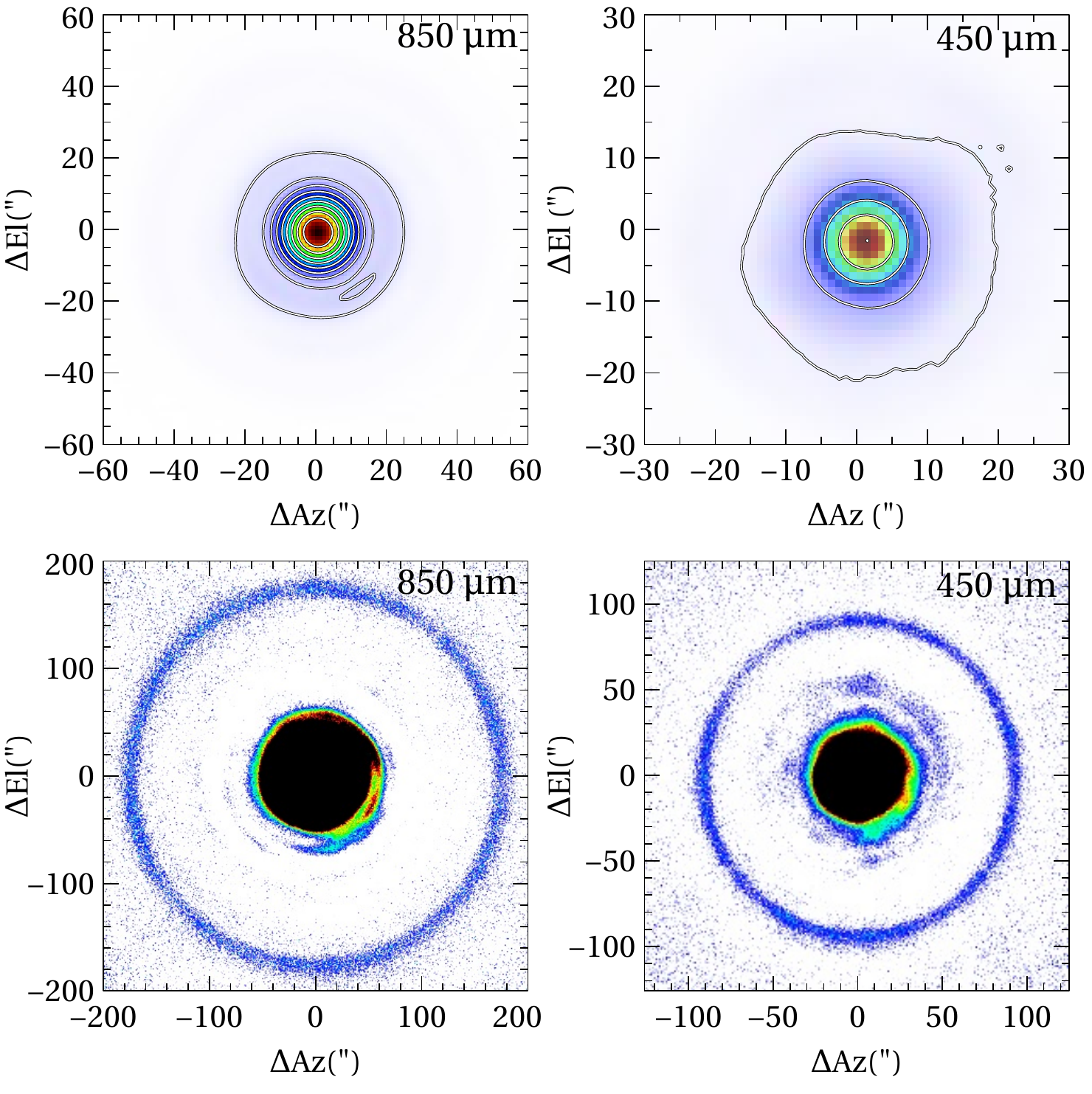}
   \caption
   {The top left image shows the inner 120 arcseconds of the co-added map of Uranus at 850\mum, with contours in square-root scaling from the peak down to 5 per cent of the peak flux. The top right image shows the inner 60 arcseconds of the corresponding Uranus map at 450\mum. The lower plots show the low-level, large scale beam pattern at both wavelengths, with the colour scaled from 0.1 per cent of the peak flux down to zero at both wavelengths.  \label{fig:beams}
 }
 \end{center}
   \end{figure}

\begin{figure}
 \begin{center}
   \includegraphics[height=8cm]{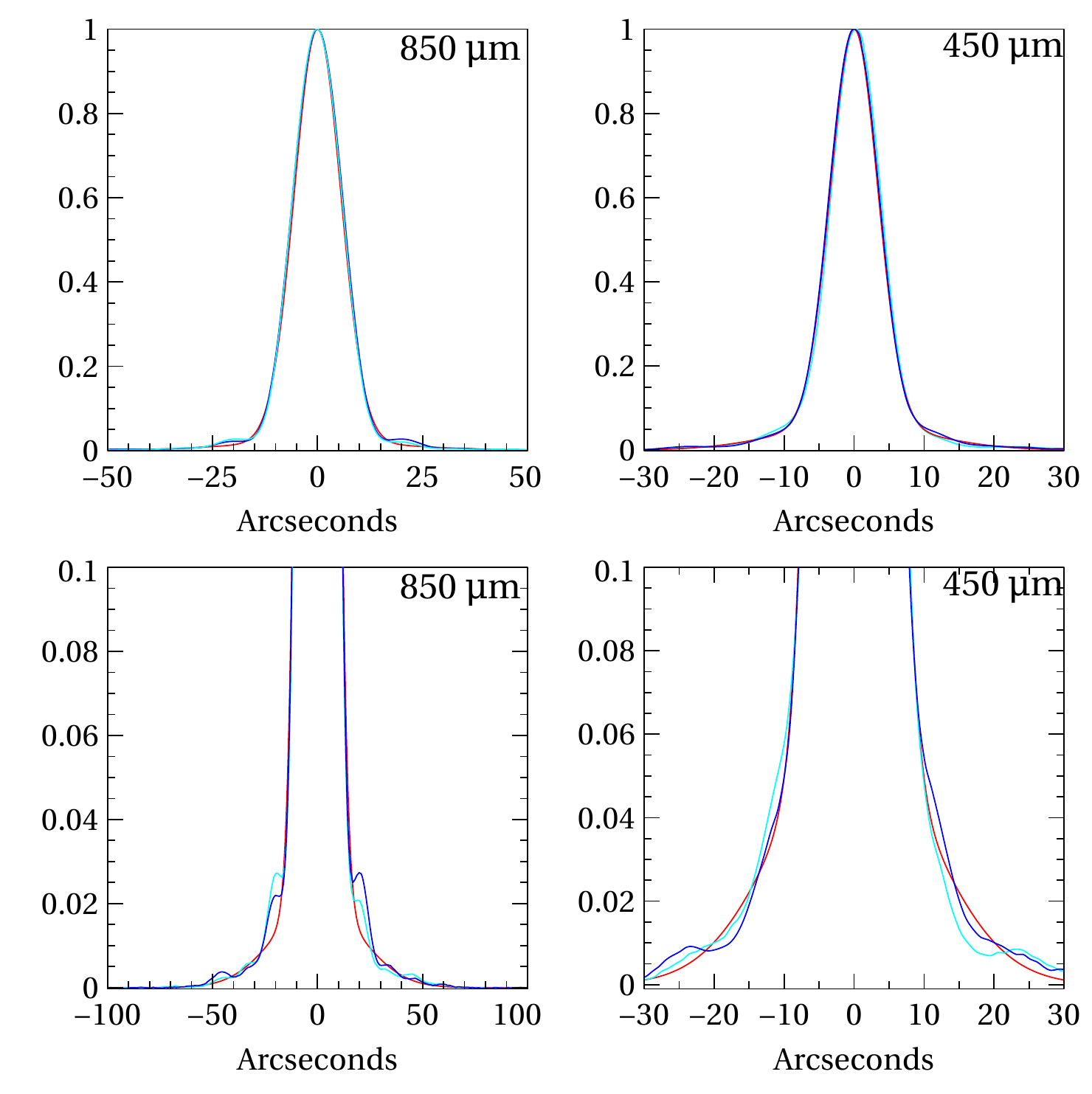}
   \caption
   { The upper plots show the azimuth (cyan) and elevation (blue) cuts through the centre of the Uranus beams from Figure~\ref{fig:beams} with the two-component fitted model overlaid in red. The lower panels show a zoom in to the secondary component of the beam. The slice through the 850\mum\ in particular, shows distinct asymmetry that is not well approximated by the Gaussian model. \label{fig:beamcuts}
 }
 \end{center}
   \end{figure}

The JCMT beam as seen with SCUBA-2 has been determined from good quality maps of Uranus obtained between 2011 May and 2012 June. The maps were reduced using the method described later in Section~\ref{obs}. The maps co-add the measured signal from all working bolometers. Individual bolometer scans of Uranus show that the beam profile varies from the centre to the edge of the focal-plane by less than 5 per cent. The individual maps were then co-added after adjustment for pointing differences. The size of Uranus was $\sim\,$3.5$\,$arcsec during this period. This reduction filters map structures on scales approximately larger than 5$\,$arcmin (\cite{chapin}). Thus any beam contribution larger than this will not be seen in SCUBA-2 maps.\\

The beam is described here in terms of two Gaussian components - a main-beam Gaussian, G$_{\rmn{MB}}$, and an additional error-beam component, G$_{\mbox{\tiny S}}$ describing the response due to large scale surface distortion and diffraction (Airy pattern). While the main beam is well described by a Gaussian this does not apply to the second component. However, using a Gaussian is a simple way to capture the volume of this error-beam. A more complete description of the second component would by asymmetric and stationary in azimuth and elevation as a result of the mounting of the telescope. It is also well known that the point source efficiency is affected by thermal surface distortions early in the evening and mid-morning. The associated aberrations would increase and possibly extend the second component. Drops of aperture efficiency as large as 50 per cent have been reported in the 450\mum\ heterodyne systems, and SCUBA measured FCF variations of up to 30 per cent at 850\mum\ under such conditions \citep{jenness2002}.\\

Figure~\ref{fig:beams} shows the co-added Uranus maps at both wavelengths. The upper plots show the inner detail of the beam pattern, while the lower plots show the low-level emission at larger scales. Normalised radial slices through the beam in the azimuth (blue) and elevation (cyan) directions are shown in Figure~\ref{fig:beamcuts}. More distinct asymmetry is noted in the 850\mum\ cuts than at 450\mum. The beam shape was then fitted as the normalised sum of the two Gaussian components as given by:

\begin{equation}
{\rmn{G}}_{\rmn{total}} = \alpha \;{\rmn{G}}_{\rmn{MB}} + \beta \; {\rmn{G}}_{\rmn{S}},
\end{equation}
where the Gaussian profile, G, is described as $exp[-4\ln2 ({r}/{\uptheta})^2]$ where $r$ is the radial distance from the centre in arcseconds and $\uptheta$ is the full-width half maximum (FWHM) in arcsec. The result of the Gaussian fitting is given in Table \ref{tab:gauss}. The integral of the profile provides the volume,

\begin{equation}
 \rmn{V}=\frac{\pi}{4 ln(2)}(\alpha(\uptheta_{\rmn {MB}})^2 + \beta(\uptheta_{\rmn{S}})^2),
\end{equation}
in arcsec$^{2}$. The widening effect resulting from the finite size of Uranus has been removed from the widths given in the table. The main beam widths are close to theoretical values estimated with a 3 dB taper - these are 7.4 and 12.75$\,$arcsec, respectively. This includes widening caused by the finite pixel size F$/2\lambda$ at 850\mum\ and F$/\lambda$ at 450\mum.\\

Apart from not estimating the taper correctly the measured beam is also affected by the atmosphere (seeing), telescope focusing errors as well as large scale surface errors. In addition slight pointing shifts between the co-added maps will also widen the beam even if the maps are aligned with each other before co-adding. Despite these factors, the volume of the error-beam is satisfactorily described by this model. Subtracting the modeled main-beam Gaussian from the co-added Uranus map in Figure~\ref{fig:beams} and integrating over the remaining volume allows empirical comparison of the error-beam with that described by the model. The measured error-beam is 24 and 39 per cent of the total integrated flux at 850 and 450\mum, respectively, compared to 25 and 40 per cent predicted by the model. The size of the second Gaussian is rather close to the first maximum in the Airy pattern from a uniform aperture. The maxima are roughly at diameters 40 and 20 arcseconds for 850 and 450\mum, respectively. \\

Strong point sources observations clearly show a low-level emission ring at large scales (see Figure~\ref{fig:beams}). The feature results from a slight mis-match between the telescope focal length and the focal length of the individual panels. The ring is 350 and 183$\arcsec$ in diameter at 850 and 450\mum, respectively. Though visible, the contribution to the beam integral is less than 1 per cent (850\mum) and 3 percent (450\mum). This contribution is not included in Table~\ref{tab:gauss}.\\

\begin{table}
\caption{Results of two component Gaussian fits\label{tab:gauss}}
\centering 
\begin{tabular}{l c c} 
\hline
  & 450 & 850 \\ [0.5ex] 
\hline 
FWHM Main beam ($\uptheta_{\rmn {MB}}$) & 7.9$\arcsec$ & 13.0$\arcsec$ \\
FWHM Secondary Comp. ($\uptheta_{\rmn {S}}$) & 25$\arcsec$ & 48$\arcsec$ \\
Rel. Amp. Main Beam ($\alpha$) & 0.94 & 0.98 \\
Rel. Amp. Secondary Comp ($\beta$) & 0.06 & 0.02 \\
Rel. Vol. Main & 0.6 & 0.75 \\
Rel. Vol. Secondary  Comp & 0.4 & 0.25 \\ [1ex] 
\hline 
\end{tabular}
\end{table}

\section[]{Flux Calibration}

Nightly flux calibration is achieved by observation of astronomical sources with known flux properties. To convert the measured signal of a source from picowatts to janskys a flux conversion factor must be applied. The primary factors affecting the FCF are the optical path to the detectors, including the shape of the dish, and the filter profiles at each wavelength.  If the optical quality is stable, and flat-fielding of the instrument and extinction correction are properly calculated, the FCF should be a constant factor. 

\subsection[]{Calibration sources}\label{calsources}
Mars and Uranus were the primary calibrators. The brightness temperatures of these planets were obtained using the JCMT {\sc fluxes} software \citep{privett} based on models from \cite{wright1976} and Moreno (2010)\footnote{Moreno, R. ``Neptune and Uranus planetary brightness temperature tabulation'' ESA Herschel Science Centre, ftp://ftp.sciops.esa.int/pub/hsc-calibration, 2010} respectively. It should be noted that the newer Uranus brightness temperatures differ from those used for SCUBA (a decrease of approximately 5 per cent and 2.5 per cent at 850\mum\ and 450\mum\ respectively). \cite{wright1976} estimates an uncertainty in the Mars model flux of $\pm$5 per cent. The Uranus model is quoted as having an absolute uncertainty of 5 per cent. Secondary calibrators used on a nightly basis were CRL$\;$618 and CRL$\;$2688. A set of additional potential secondary calibrators were also observed and are listed with their SCUBA fluxes in Table~\ref{tab:cals}.

\begin{table*}
  \begin{minipage}{160mm}
\begin{footnotesize}
\caption{Previously published flux densities of SCUBA-2 potential secondary calibrators.\label{tab:cals} }
 \begin{tabular}{lccc|cccc}
  \toprule
\multicolumn{1}{c}{SOURCE}             &
\multicolumn{1}{c}{RA}             &
\multicolumn{1}{c}{Dec.}             &
\multicolumn{1}{c}{850 [I]}          &
\multicolumn{1}{c}{450 [I]}          &
\multicolumn{1}{c}{850 [P]}          &
\multicolumn{1}{c}{450 [P]}          &
\multicolumn{1}{c}{ref}          \\

\multicolumn{1}{c}{}             &
\multicolumn{1}{c}{(J2000)}             &
\multicolumn{1}{c}{(J2000)}             &
\multicolumn{1}{c}{(Jy)}          &
\multicolumn{1}{c}{(Jy)}          &
\multicolumn{1}{c}{(Jy beam$^{-1}$)}          &
\multicolumn{1}{c}{(Jy beam$^{-1}$)}          &
\multicolumn{1}{c}{}          \\
\midrule
CRL618 & 04:42:53.67& +36:06:53.17&4.73 $\pm$ 0.33 & 12.1 $\pm$ 2.2 & 4.55 $\pm$ 0.2 & 11.5 $\pm$ 1.5 &d$^{[I]}$,b$^{[P]}$\\  
CRL2688&  21:02:18.27& +36:41:37.00&6.39 $\pm$ 0.51 & 30.9 $\pm$ 3.8 & 5.9 $\pm$ 0.2 & 24.0 $\pm$ 2.1 &d$^{[I]}$,b$^{[P]}$\\
Arp220&  15:34:57.27& +23:30:10.48& 0.83 $\pm$ 0.086 & --- &0.7$^*$ & 5.0$^*$ &f$^{[I]}$,c$^{[P]}$ \\
PV Cep&  20:45:54.39& +67:57:38.8& 1.0 $\pm$ 0.02 & 6.5  $\pm$ 0.015 & --- & --- &a\\
MWC 349& 20:32:45.53& +40:39:36.6&2.6 $\pm$ 0.007 & 5 $\pm$ 1.1 & --- & --- &a  \\
HD169142& 18:24:29.78& -29:46:49.37& 0.565 $\pm$ 0.01& 3.34 $\pm$ 0.115 & ---& ---&a \\
V883 Ori&05:38:19 & -07:02:2&1.41 $\pm$ 0.022 & 9.45 $\pm$ 0.17 & --- & --- & g \\
HL Tau&04:31:38.44 & +18:13:57.65 & 2.36 $\pm$ 0.24 & 9.9 $\pm$ 2.0 & 2.3 $\pm$ 0.15 & 10.0 $\pm$ 1.2 &d$^{[I]}$,b$^{[P]}$\\
BVP 1 & 17:43:10.37 & -29:51:44.00 &  --- & --- & 1.069 $\pm$ 0.02  & 11.318 $\pm$ 0.6 & e\\
\bottomrule
\end{tabular}
\medskip

[I] indicates integrated flux  measurements. [P] indicates peak flux measurements. References: (a) \citet{sandell2011}, background-corrected integrated flux. (b) \citet{sandell2003}, peak flux in Jy beam$^{-1}$. (c) \citet{truch}, BLAST primary calibrator. $^*$Peak flux based on fit to SED curve fitted to measurements at other wavelengths.(d) \citet{jenness2002} integrated flux in a 40 arcsec aperture. (e) \citet{barnard} Peak flux. Denoted as object G6 in that text. (f) \citet{lisenfeld} integrated flux. (g) \citet{sandell2001}, background-corrected integrated flux.
\end{footnotesize}
 \end{minipage}

\end{table*}

\subsection{Observation method and data reduction}\label{obs}

All calibrators were observed using a daisy scan-pattern \citep{holland2012}, designed specifically for observation of point and compact sources, with observations lasting approximately four minutes each. The data were reduced using the SMURF routine {\sc makemap} \citep{chapin,jenness2011}, using a parameter file designed specifically for bright point sources. The reduction of compact calibrators uses a combination of common-mode rejection and high-pass filtering of the bolometer data (which effectively suppresses information on scales larger than 3.3 arcmin) to remove low-frequency noise. The ringing one might expect from such filtering is then controlled by constraining the map to a value of zero beyond 60\,arcsec from the source (well beyond the bulk of the power in the central lobe) for all but the final iteration. For point-like sources, the gain (independent of the brightness of the source) is effectively one. These parameters are discussed in detail in Section 4.1 of \cite{chapin}. The calibrations were taken between 2011 May and 2012 May, and over 500 observations (at each wavelength) were included in this analysis. Maps were regridded to a 1-arcsec pixel scale, to optimise the quality of the fits to the peak of the beam.\\

\subsection{Aperture photometry}\label{apphotom}
\begin{figure}
   \includegraphics[height=5.0cm]{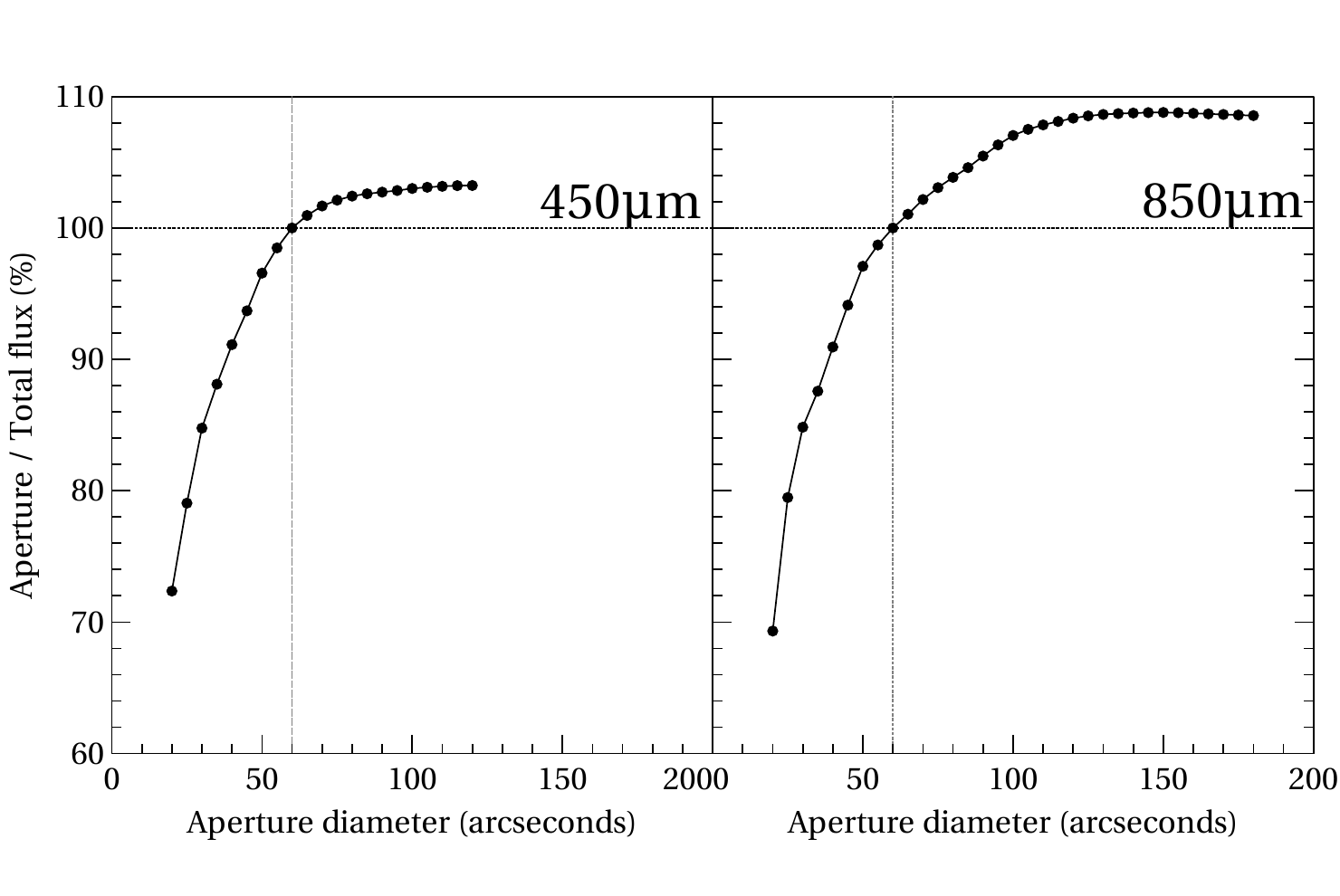}
   \caption[cog]
   { \label{fig:cog}
Aperture photometry curve of growth normalised for a 60 arcsec aperture, at both wavelengths.}
   \end{figure}

The total integrated flux for each calibrator map was calculated in a 60$\,$arcsec diameter aperture, with the background level calculated within an annulus of inner diameter 90$\,$arcsec and outer diameter 120$\,$arcsec (1.5 and 2.0 times the source aperture). As discussed in Section~\ref{beam}, the SCUBA-2 beam has a broad error-beam, and a 60$\,$arcsec aperture does not fully encompass the beam, particularly at 850\mum. However, increasing the aperture, and thus by necessity, the annulus, beyond this size showed a significant increase in noise, and produced larger scatter in the results.\\

The 60$\,$arcsec aperture was chosen as a compromise between minimising the scatter in the results whilst optimising the total flux measurement. The curve of growth in Figure~\ref{fig:cog} shows the ratio of Aperture flux to Total Flux for a co-add of several Uranus observations, normalised to an aperture of 60$\,$arcsec. At 450\mum, this aperture underestimates the total flux by approximately 4 per cent while at 850\mum\ this value is approximately 8 per cent. In the results presented here, the Uranus calibration maps and the secondary sources are both reduced using a 60$\,$arcsec aperture, therefore no correction is required in the FCF values quoted below. However, if a different aperture size is chosen for a science source, the result needs to be {\it divided} by the appropriate factor derived from Figure~\ref{fig:cog} and listed for reference in the lookup Table~\ref{tab:cog}. These factors are only applicable to photometry of unresolved sources.\\

\begin{table}
\caption{When integrating over an aperture of diameter, d, in arcseconds, the FCF given in Section~\ref{results} must be divided by the following factor, given in columns denoted 450$_F$ and 850$_F$. The factors are derived from the measurements plotted in Figure~\ref{fig:cog}. These factors are only valid for unresolved sources.\label{tab:cog}}
\centering 
\begin{tabular}{l l l} 
\hline
$d$ (arcsec)  & 450$_F$  & 850$_F$  \\ [0.5ex] 
\hline 
20 &  0.72  &  0.69 \\
25 & 0.79    & 0.79 \\
30 & 0.85    & 0.85 \\
35 & 0.88    & 0.88 \\
40 & 0.91    & 0.91 \\
45 &0.94     &0.94 \\
50 &0.97     &0.97 \\
55 &0.98     &0.99 \\
60 &1.00     &1.00 \\
65 &1.01     &1.01 \\
70 &1.02     &1.02 \\
75 &1.02     &1.03 \\
80 &1.02     &1.04 \\
85 &1.03     &1.05 \\
90 &1.03     &1.05 \\
95 &1.03     &1.06 \\
100& 1.03    & 1.07 \\
105$-$120  & 1.03    &   1.08  \\
$>$120 & --- & 1.09  \\
\hline
\end{tabular}
\end{table}

The FCF is simply described as the ratio between the known flux  $S$, in Jy, and the measured signal in pW, $I_0$, summed over the source. To apply the FCF when calculating a flux integrated over a source, the size of the pixels in the output map must also be taken into account. This factor is called the FCF$_{\rmn {arcsec}}$, and is given by:
\begin{equation}
{\rmn {FCF}}_{\rmn {arcsec}} = \frac{S}{I_0 \; A},
\label{fcfeq1}
\end{equation}
where $A$ is the pixel area in arcsec$^2$, giving the units of  Jy$\;$pW$^{-1}\;$arcsec$^{-2}$. \\

It is sometimes preferable to measure the peak flux of a source, rather than the integrated value. To fit the peak of each calibration map, a Gaussian-like 2D fit was applied to each calibrator using the KAPPA application {\sc beamfit} \citep{currie}, with the power index allowed to be a variable in the fit. Unlike the total flux calculated in a large aperture, the source peak is highly susceptible to errors in telescope focus, as well as temperature-dependent distortions in the dish shape, most noticeable early in evening observations. The peak FCF, FCF$_{\rmn {peak}}$, is simply the ratio of the known peak flux in Jy to the measured peak flux in pW. 

\subsection{Results}\label{results}
\begin{figure}
   \includegraphics[height=8.4cm]{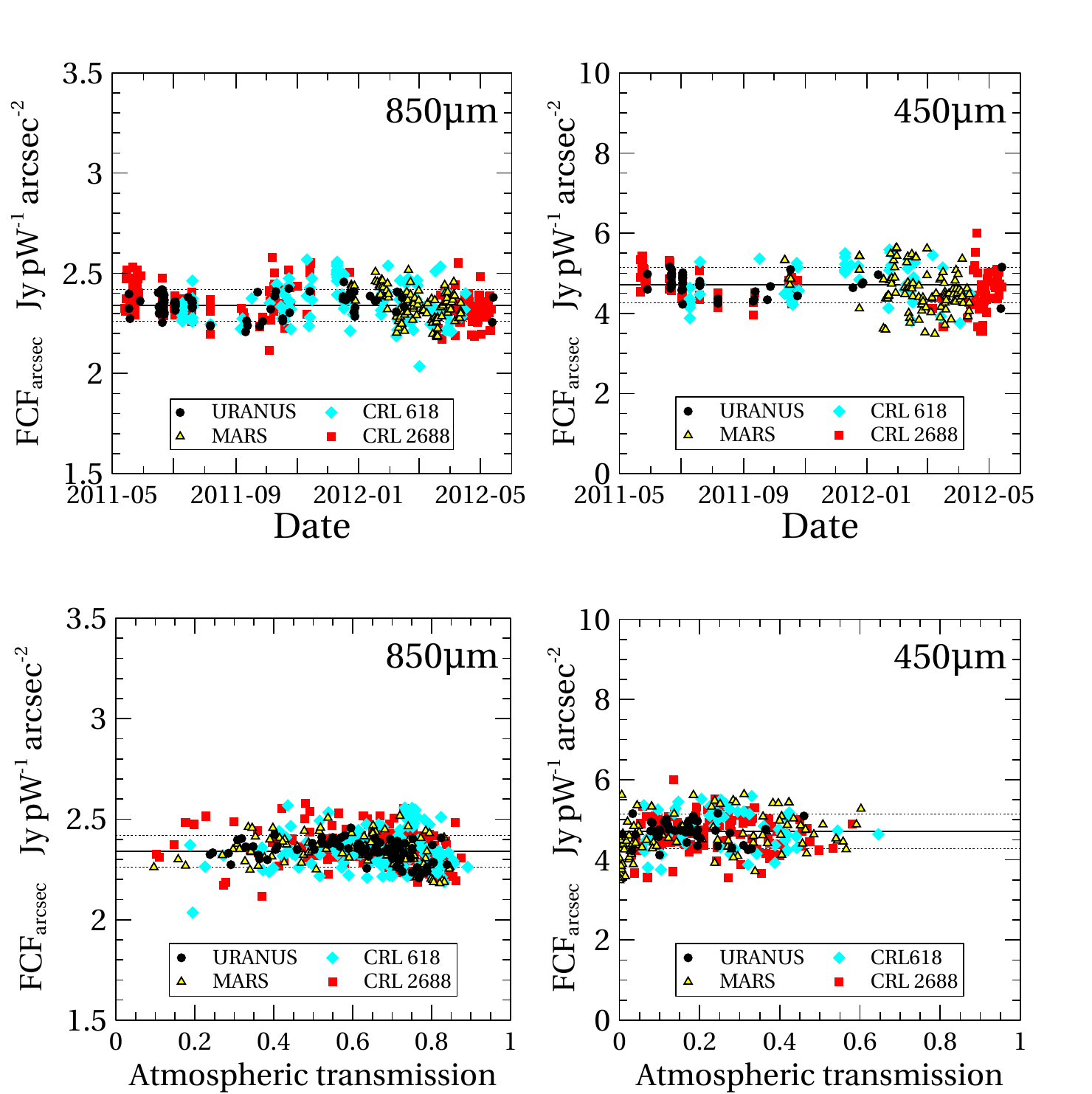}
   \caption[850fcf]
   { \label{fig:fcfasec}
Flux conversion factor, FCF$_{\rmn {arcsec}}$ results, at 850\mum\ (left) and 450\mum\ (right), plotted as a function of time (top) between May 2011 and May 2012, and as a function of atmospheric transmission (bottom). The solid line indicates the average value, and the dotted lines denote the 1-$\sigma$ error.}
   \end{figure}

\begin{figure}
   \includegraphics[height=8.4cm]{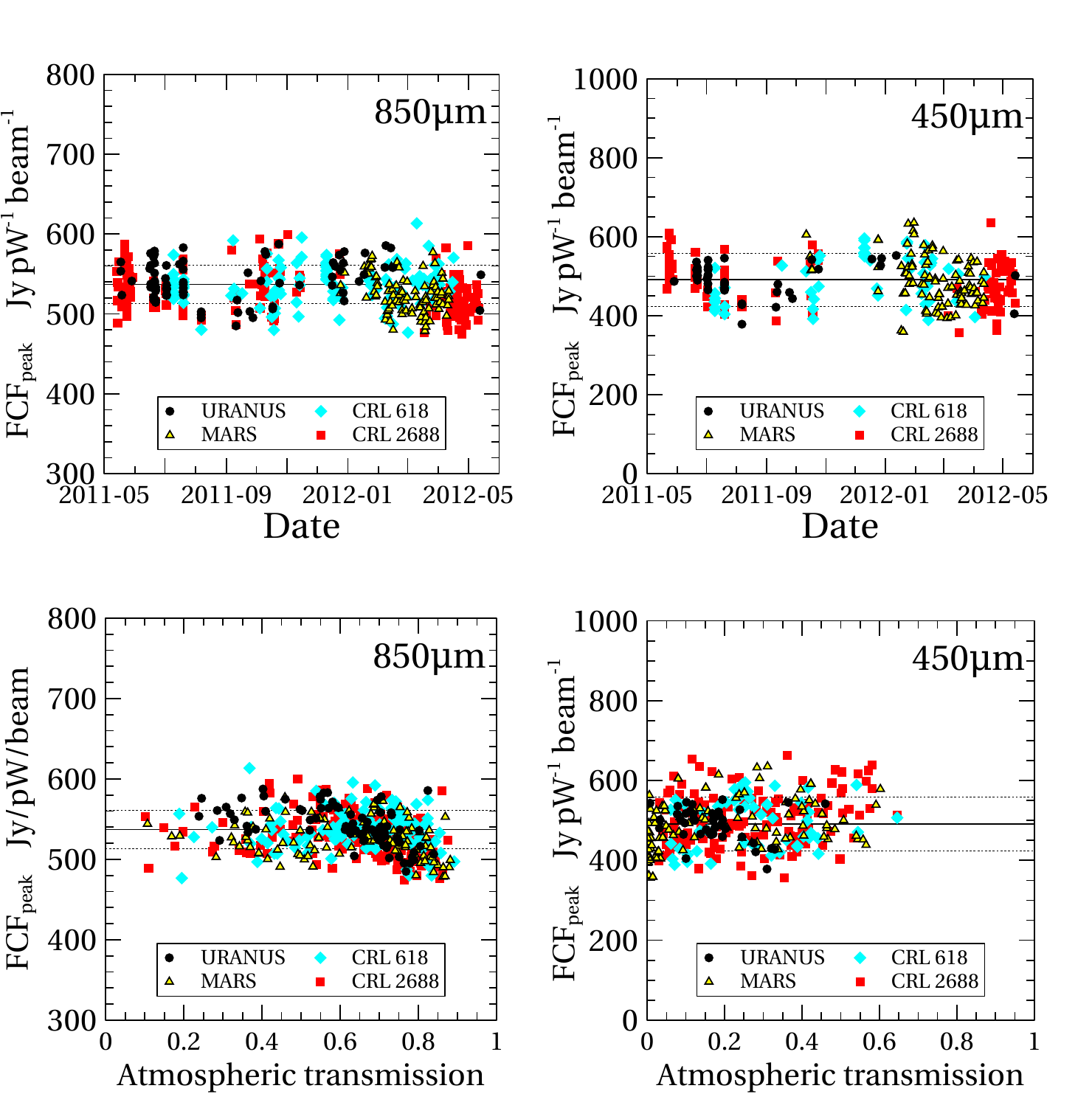}
   \caption[450fcf]
   { \label{fig:fcfpeak}
Flux conversion factor, FCF$_{\rmn {peak}}$ results, at 850\mum\ (left) and 450\mum\ (right), plotted as a function of time (top) between May 2011 and May 2012, and as a function of atmospheric transmission (bottom). The solid line indicates the average value, and the dotted lines denote the 1-$\sigma$ error.}
   \end{figure}

\begin{figure}
   \includegraphics[height=8.4cm]{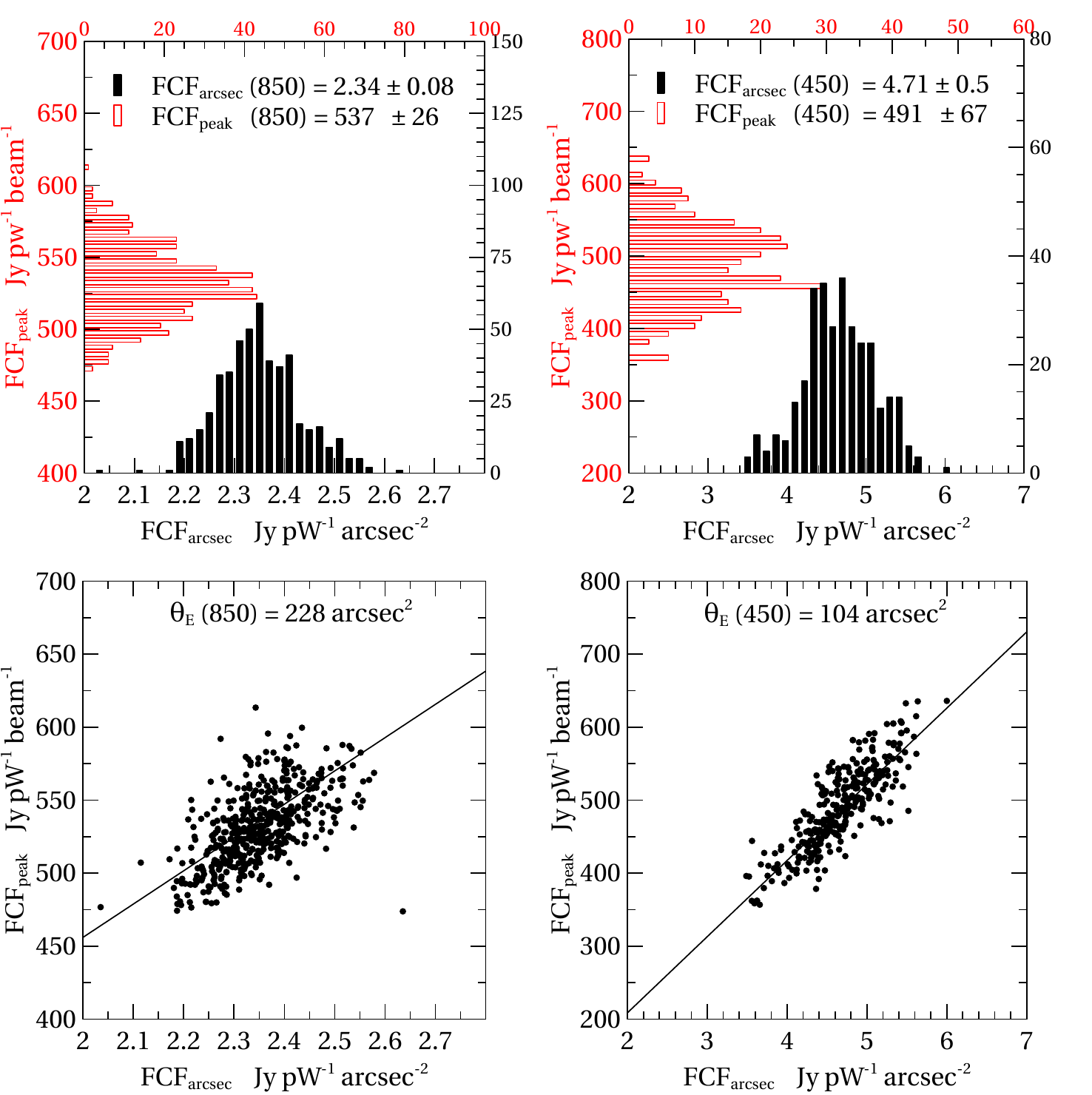}
   \caption[450fcf]
   { \label{fig:fcfhist}
The top left plot shows the histogram distribution of the FCF$_{\rmn {arcsec}}$ (black, solid) and FCF$_{\rmn {peak}}$ (red, outline)  at 850\mum\ for the data presented in Figure~\ref{fig:fcfasec}, while the top right plot shows the corresponding distributions at 450\mum. The two lower panels show the FCF$_{\rmn {arcsec}}$ versus FCF$_{\rmn {peak}}$ at both wavelengths. A linear fit to this relation allows an empirical estimate of the beam area. }
   \end{figure}

Figures \ref{fig:fcfasec}, \ref{fig:fcfpeak} and \ref{fig:fcfhist}  show the results of the FCF calculations for the standard calibrators, Uranus, Mars, CRL618 and CRL2688. Figure~\ref{fig:fcfasec} shows FCF$_{\rmn {arcsec}}$ as a function of time (top), and versus atmospheric transmission (bottom). At both wavelengths the results show no systematic deviation, indicating the instrument and optical performance were consistent during this year-long period.\\

Figure~\ref{fig:fcfpeak} shows the same comparisons, but for the FCF$_{\rmn {peak}}$. These results shows larger deviation from the mean, particularly at 450\mum. This is to be expected given this measure is more susceptible to focus and temperature variations in the optics, however, again, no systematic deviation is observed.  Figure~\ref{fig:fcfhist} shows the histogram distribution of FCF$_{\rmn {arcsec}}$ and FCF$_{\rmn {peak}}$  for the combined set of calibrations at 850\mum\ (top left panel) and 450\mum\ (top right panel). The resulting mean FCFs at each wavelength are thus:
\begin{equation}
{\rmn {FCF[850]}}_{\rmn {arcsec}} = 2.34 \pm 0.08 \;\;{\rmn {Jy\;pW^{-1}\; arcsec^{-2}}}
\end{equation}

and
\begin{equation}
{\rmn {FCF[450]}}_{\rmn {arcsec}} = 4.71 \pm 0.5  \;\;{\rmn {Jy\;pW^{-1}\;arcsec^{-2}}}.
\end{equation}

The corresponding peak FCF results are:
\begin{equation}
{\rmn{FCF[850]}}_{\rmn {peak}} = 537 \pm 26 \;\;{\rmn {Jy\;pW^{-1}}}
\end{equation}

and
\begin{equation}
{\rmn{FCF[450]}}_{\rmn{peak}} = 491 \pm 67  \;\;{\rmn {Jy\;pW^{-1}}}.
\end{equation}
The error is the 1-$\sigma$ deviation from the mean. The relative flux calibration is thus better than 5 per cent at 850\mum\ and $\sim$ 10 per cent at 450\mum. In comparison, the relative flux calibration of SCUBA from \cite{jenness2002} was 10 per cent at 850\mum\ and 20 per cent at 450\mum. Current instruments with comparable wavebands include LABOCA, on the APEX telescope \citep{siringo} which measures a calibration accuracy of 10 per cent at 870\mum\ and BOLOCAM, at the Caltech Submillimeter Observatory, with an 8 per cent uncertainty at 1.1 millimetres \citep{aguirre2011}. SHARC-II, the 350\mum\ bolometer camera at Caltech Submillimeter Observatory, reports a calibration uncertainty of 15 per cent \citep{kovacs}, and as it shares the same site as SCUBA-2 it provides a point of comparison for the SCUBA-2 450\mum\ accuracy, as the 350\mum\ and 450\mum\  atmospheric windows are similarly affected by small variations in PWV. This improvement by a factor of two in the calibration accuracy at both wavelengths, in comparison to SCUBA, can be attributed to the increased time-resolution measurements of the atmospheric conditions by the WVM, and the stability of the instrument performance.\\

The absolute calibration accuracy must also take into account the uncertainty in the modeled fluxes of our primary calibrators, Mars and Uranus, which as noted in Section~\ref{calsources} are both $\pm$ 5 per cent. Less than a percent difference is observed in the mean FCF derived for the calibrators using only one or other of these sources implying that the estimate of the uncertainty on these models is an upper limit and we have a better calibration than the addition of this error estimate would suggest. Taking the case that we have an overall model error of 5 per cent, the quadrature addition of the model error and relative FCF error would give an upper limit on the absolute calibration uncertainty of 8 per cent at 850\mum\ and 12 per cent at 450\mum. \\

The lower panels in Figure~\ref{fig:fcfhist} compare the FCF$_{\rmn {peak}}$ versus FCF$_{\rmn {arcsec}}$ at each wavelength. This allows an empirical estimation of the beam area, $A_{\rmn {E}}$, by fitting the linear relation between the peak and integrated values. The beam areas derived in this way therefore determine the effective FWHM at each wavelength. Assuming a gaussian beam, the effective FWHM, $\uptheta_{\rmn {E}}$, at each wavelength can be determined as $\sqrt{{A_{\rmn {E}}}/1.133)}$, resulting in $\uptheta_{\rmn {E}}$ of 14.1 and 9.6$\,$arcsec at 850 and 450\mum\ respectively. Note that this is wider than the fit to the main beam component at each wavelength as shown in Table~\ref{tab:gauss}, as it accounts for the extended emission in the shoulders of the beam. The effective FWHM should relate to the derived, two-component beam as $\uptheta_{\rmn {E}} = \sqrt{\alpha (\uptheta_{\rmn {MB}})^2 + \beta (\uptheta_{\rmn{S}})^2}$. The $\uptheta_{\rmn {E}}$ predicted from the two-component model gives 14.6 and 9.8$\,$arcsec at 850 and 450\mum\ respectively, which compares reasonably with the estimate derived from the calibrations.\\

\section[]{Calibration source fluxes}

\begin{table*}
  \begin{minipage}{160mm}
\caption{Measured fluxes of SCUBA-2 secondary calibrators\label{tab:newcals}}
\begin{footnotesize}
 \begin{tabular}{lccc|ccccl}
  \toprule
\multicolumn{1}{c}{SOURCE}             &
\multicolumn{1}{c}{RA}             &
\multicolumn{1}{c}{Dec.}             &
\multicolumn{1}{c}{N(obs)}           &
 \multicolumn{1}{c}{850[I]}          &
\multicolumn{1}{c}{450[I]}          &
\multicolumn{1}{c}{850[P]}          &
\multicolumn{1}{c}{450[P]}          &
\multicolumn{1}{c}{Note}          \\

\multicolumn{1}{c}{}             &
\multicolumn{1}{c}{(J2000)}             &
\multicolumn{1}{c}{(J2000)}             &
\multicolumn{1}{c}{}           &
 \multicolumn{1}{c}{(Jy)}          &
\multicolumn{1}{c}{(Jy)}          &
\multicolumn{1}{c}{(Jy beam$^{-1}$)}          &
\multicolumn{1}{c}{(Jy beam$^{-1}$)}          &
\multicolumn{1}{c}{}          \\
\midrule
CRL618 & 04:42:53.67& +36:06:53.17&113&5.0 $\pm$ 0.2 &12.10 $\pm$ 1.05 &4.89 $\pm$ 0.24 &11.50 $\pm$ 1.4& y\\  
CRL2688&  21:02:18.27& +36:41:37.00&173&6.13 $\pm$ 0.211 &29.10 $\pm$ 2.5 &5.64 $\pm$ 0.27 &24.9 $\pm$ 2.9 &y,(a) \\
Arp220&  15:34:57.27& +23:30:10.48&20&0.81 $\pm$ 0.07 & 5.4 $\pm$ 0.7 & 0.79 $\pm$ 0.9 & 5.2 $\pm$ 0.8 & y\\
PV Cep&  20:45:54.39& +67:57:38.8&21 & 1.35 $\pm$ 0.05 &10.6 $\pm$ 0.86 & 0.82 $\pm$ 0.04 & 5.7  $\pm$ 0.65 &n,(a)\\
MWC 349& 20:32:45.53& +40:39:36.6&14 & 2.19 $\pm$ 0.08 & 3.2 $\pm$ 0.25 &2.21 $\pm$ 0.11 & 3.4 $\pm$ 0.26 & n\\
HD169142& 18:24:29.78& -29:46:49.37&12& 0.58 $\pm$ 0.02 & 3.41 $\pm$ 0.24 & 0.52 $\pm$ 0.03 & 2.21 $\pm$ 0.25 & y, (a,b)\\
V883 Ori&05:38:19 & -07:02:2&3& 2.00 $\pm$ 0.07 & 10.4 $\pm$ 1.0 &1.55 $\pm$  0.09 & 7.8 $\pm$ 1.00 & y, (b)\\
HL Tau&04:31:38.44 &+ 18:13:57.65 & 9& 2.42 $\pm$ 0.08 & 10.3 $\pm$ 0.86 & 2.32 $\pm$ 0.1 & 8.3 $\pm$ 1.03 &y, (b)\\
BVP 1 & 17:43:10.37 & -29:51:44.00 & 4 & 1.55 $\pm$ 0.05 &  16.9 $\pm$ 0.8 & 1.37 $\pm$ 0.07  & 11.5 $\pm$ 1.5 & y,(a,b)\\
\bottomrule
\end{tabular}
\medskip

[I] indicates integrated flux measurements. [P] indicates peak flux measurements. Error is quadrature addition of uncertainty in the FCF measurement and the rms noise in the co-added map for each calibrator. Notes: y/n indicates if the source will continue to be observed as a SCUBA-2 calibrator. (a) Extended emission at 450\mum. Col.$\,$4 gives the number of observations taken per source between 2011 July and 2012 May. (b) Low S/N measurement, particularly at 450\mum. Accuracy of these flux densities is uncertain, and further observations are required.
\end{footnotesize}
 \end{minipage}
\end{table*}

Several well-studied submillimetre point sources were also regularly observed as potential new calibrators. Using the new extinction corrections and FCFs derived above, these sources were calibrated and compared with previous determinations of their flux. Table~\ref{tab:newcals} shows the calibrator fluxes for all potential calibrators that were observed. In all cases, a co-added map of a large set (or in some cases, all) of the observations was produced and the integrated fluxes were calculated in a 60$\,$arcsec aperture as described in Section~\ref{apphotom}. The table reports the integrated fluxes and also the peak fluxes as derived from a single Gaussian fit to the source profile. 

\subsection{CRL$\,$618}
\begin{figure}
   \includegraphics[height=4.2cm]{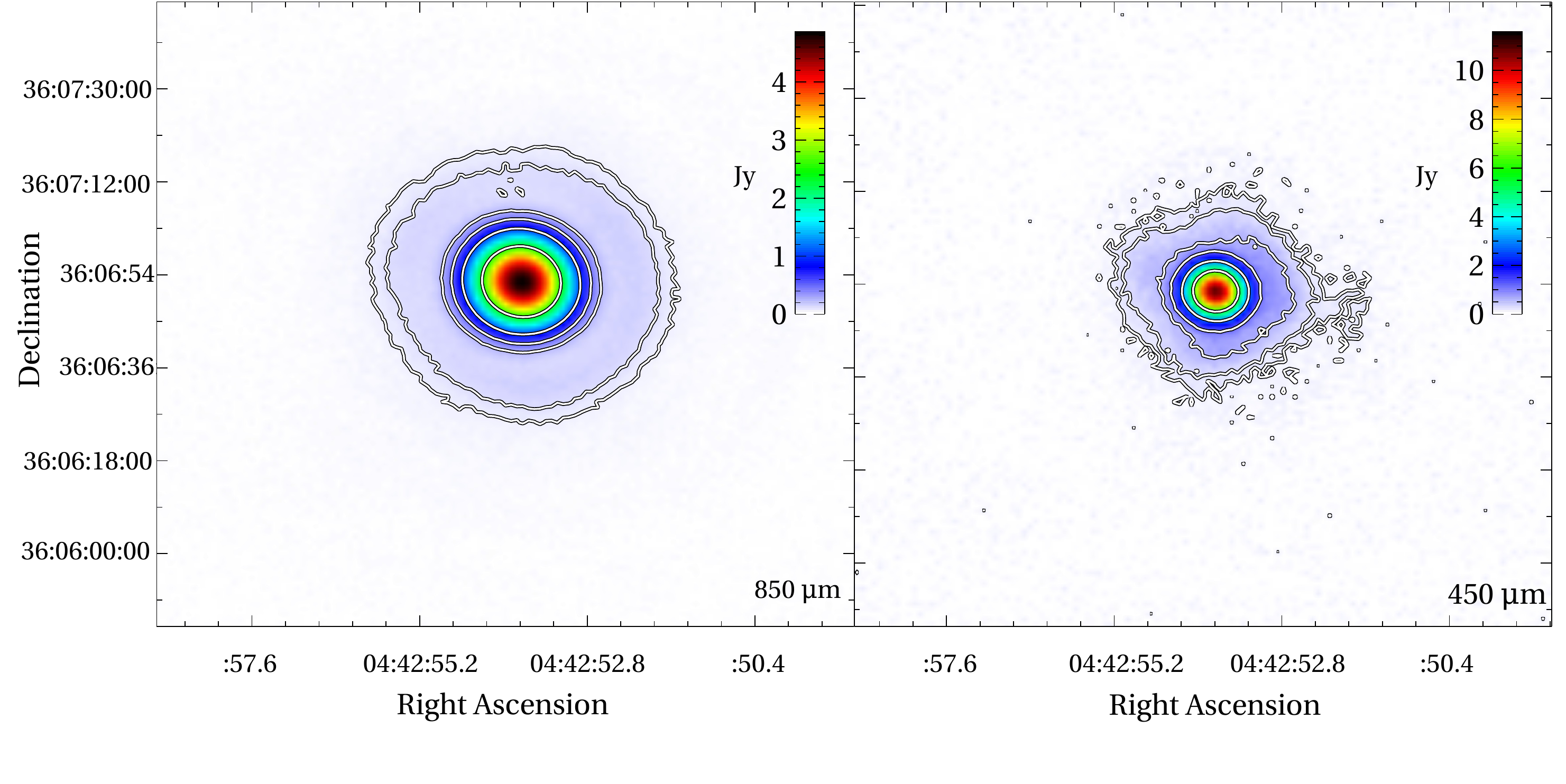}
   \caption[crl618]
   { \label{fig:crl618}
850\mum\ (left) and 450\mum\ (right) image of CRL$\,$618. The colour scales linearly from the background level to the peak as indicated by the colour bar on the right-hand side of each plot. Contours indicate  1, 2, 5, 10, 20 and 50 per cent of the peak flux. The 3-$\sigma$ level at 850\mum\ is 0.3 per cent of the peak flux at 17$\,$mJy, while at 450\mum\, 3-$\sigma$ is 1 per cent of the peak at 116$\,$mJy.}
   \end{figure}

CRL$\,$618 is also extremely well-observed in the submillimetre \citep{sandell2003,jenness2010,jenness2002,siringo} and is compact, with a measured FWHM of 13.7 and 8.0$\,$arcsec at 850 and 450\mum, respectively. Figure~\ref{fig:crl618} shows CRL$\,$618 at both wavelengths. The calculated integrated flux at 850\mum\ is slightly higher at 5.0$\,$Jy in comparison to results presented by \cite{jenness2002}, though the result does fall within their error range. The possibility that this source varies over time has been suggested \citep{knapp}. No source variability was seen in the SCUBA-2 results between 2011 July and 2012 May, which agrees with the results from SCUBA between 1997 and 2005 as presented by \cite{jenness2010}. At 870\mum\, LABOCA measure a peak flux of 4.9$\pm$0.2$\,$Jy beam$^{-1}$\citep{siringo} which agrees well with our result of 4.89$\pm$0.24$\,$Jy beam$^{-1}$. At 450\mum, the integrated and peak fluxes agree extremely well with the published SCUBA counterparts.\\

\subsection{CRL$\,$2688}

\begin{figure}
   \includegraphics[height=4.2cm]{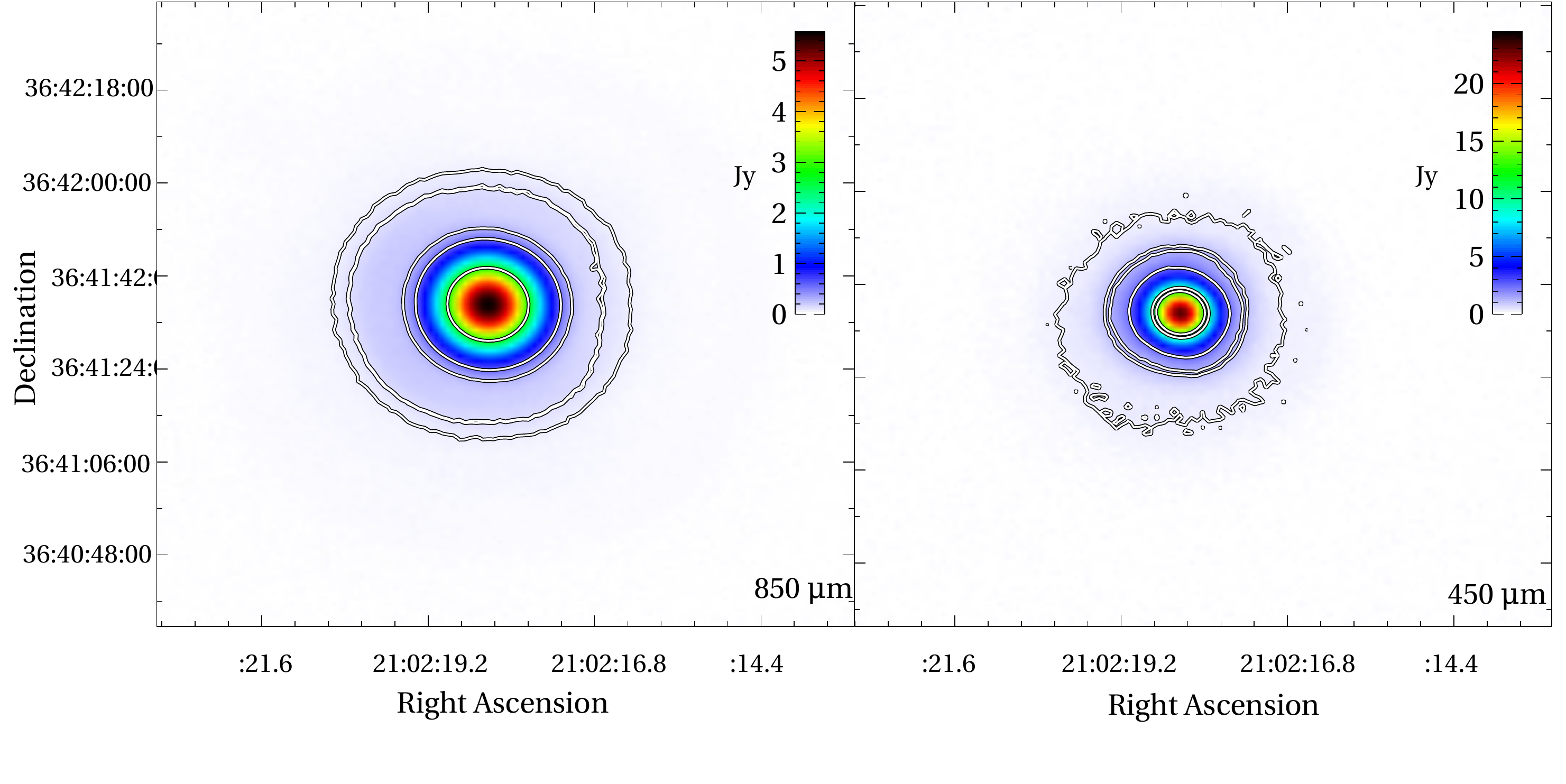}
   \caption[crl2688]
   { \label{fig:crl2688}
850\mum\ (left) and 450\mum\ (right) image of CRL$\,$2688. The colour scales linearly from the background level to the peak as indicated by the colour bar on the right-hand side of each plot. Contours indicate  1, 2, 5, 10, 20 and 50 per cent of the peak flux. The 3-$\sigma$ level at 850\mum\ is 0.2 per cent of the peak flux at 11$\,$mJy, while at 450\mum\, 3-$\sigma$ is 0.5 per cent of the peak at 83$\,$mJy.}
   \end{figure}

 CRL$\,$2688 is a well-known compact submillimetre source that has been used as a calibrator for SCUBA \citep{sandell2003}, and other submillimetre instruments including BLAST \citep{truch}. Over 170 observations of the source were taken with SCUBA-2 during 2011 and early 2012, and the co-added map of 55 of the best weather observations is shown at both wavelengths in Figure~\ref{fig:crl2688}. At 850\mum, the source is compact, though possibly slightly resolved, and the measured FWHM is 14.2$\,$arcsec. At 450\mum, we measure a FWHM of 9.4$\,$arcsec indicating the source is extended. This is in agreement with previous measurements and this extended emission is entirely contained within the 60$\,$arcsec aperture used to calculate the photometry. Fluxes are in good agreement with the measurement of the peak flux by \cite{sandell2003,siringo} and the integrated flux by \cite{jenness2002}. 

\subsection{Arp$\,$220}
\begin{figure}
   \includegraphics[height=4.2cm]{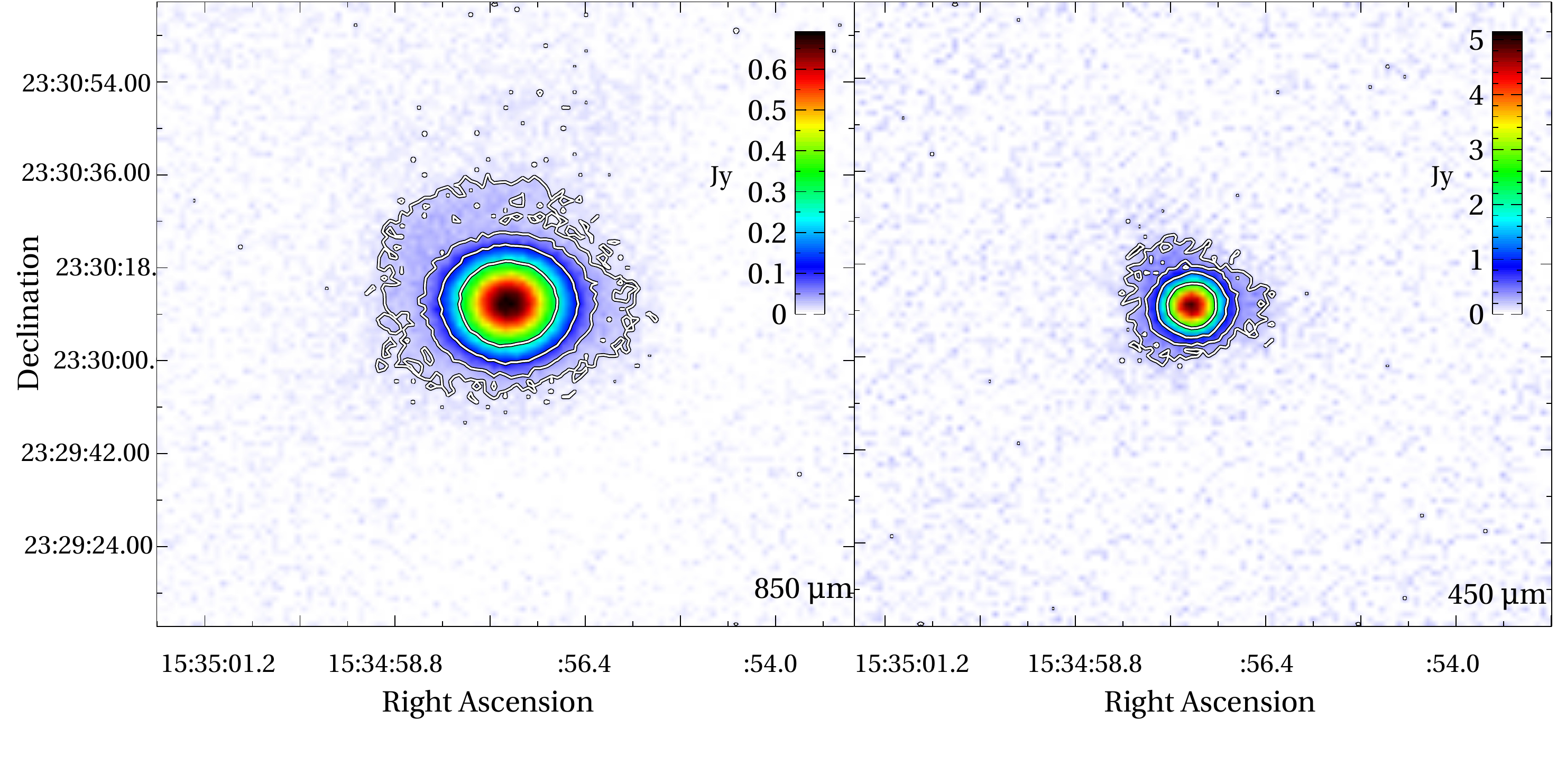}
   \caption[arp]
   { \label{fig:arp220}
850\mum\ (left) and 450\mum\ (right) image of Arp$\,$220. The colour scale is linear as shown in the colour bar on the right hand side. In each map, the five contours scale linearly from the peak signal as shown in Table~\ref{tab:newcals} down to the 3-$\sigma$ detection level (0.18/0.36$\,$Jy at 850/450\mum\ respectively). }
   \end{figure}

Arp$\,$220 is an ultra-luminous nearby galaxy and is well-studied in the submillimetre. Several publications have investigated the continuum flux at 850\mum\ \citep{dunne,lisenfeld,seiffert,sakamoto}. At 850\mum, we calculate an integrated flux of 0.81 $\pm$ 0.07$\,$Jy, compared with previous determinations ranging from 0.78$\,$Jy (\cite{sakamoto}) to 0.83$\,$Jy (\cite{dunne}). At 450\mum, the integrated flux of 5.4$\,$Jy agrees reasonably with the SED fit from \cite{truch}. There are no published continuum fluxes for this source at 450\mum. \cite{truch} fit an SED to Arp$\,$220 using a combination of published submillimetre and far-infrared flux densities, to calculate interpolated flux densities in their 250, 350, and 500\mum\ BLAST bands (the last of which overlaps significantly with the SCUBA-2 450\mum\ bandpass). The resulting fluxes derived from 20 observations of the source in 2011 June and July show good agreement with previous results. From a selection of observations in the best conditions, this unresolved source shows a measured FWHM of 13.1 arcsec at 850\mum\ and 7.95 arc sec at 450\mum. This agrees with the main-beam Gaussian fit in Table~\ref{tab:gauss}. Though faint, Arp$\,$220 will continue to be observed as a calibrator for SCUBA-2 in good weather.\\

\subsection{PV$\,$Cep}
\begin{figure}
   \includegraphics[height=4.2cm]{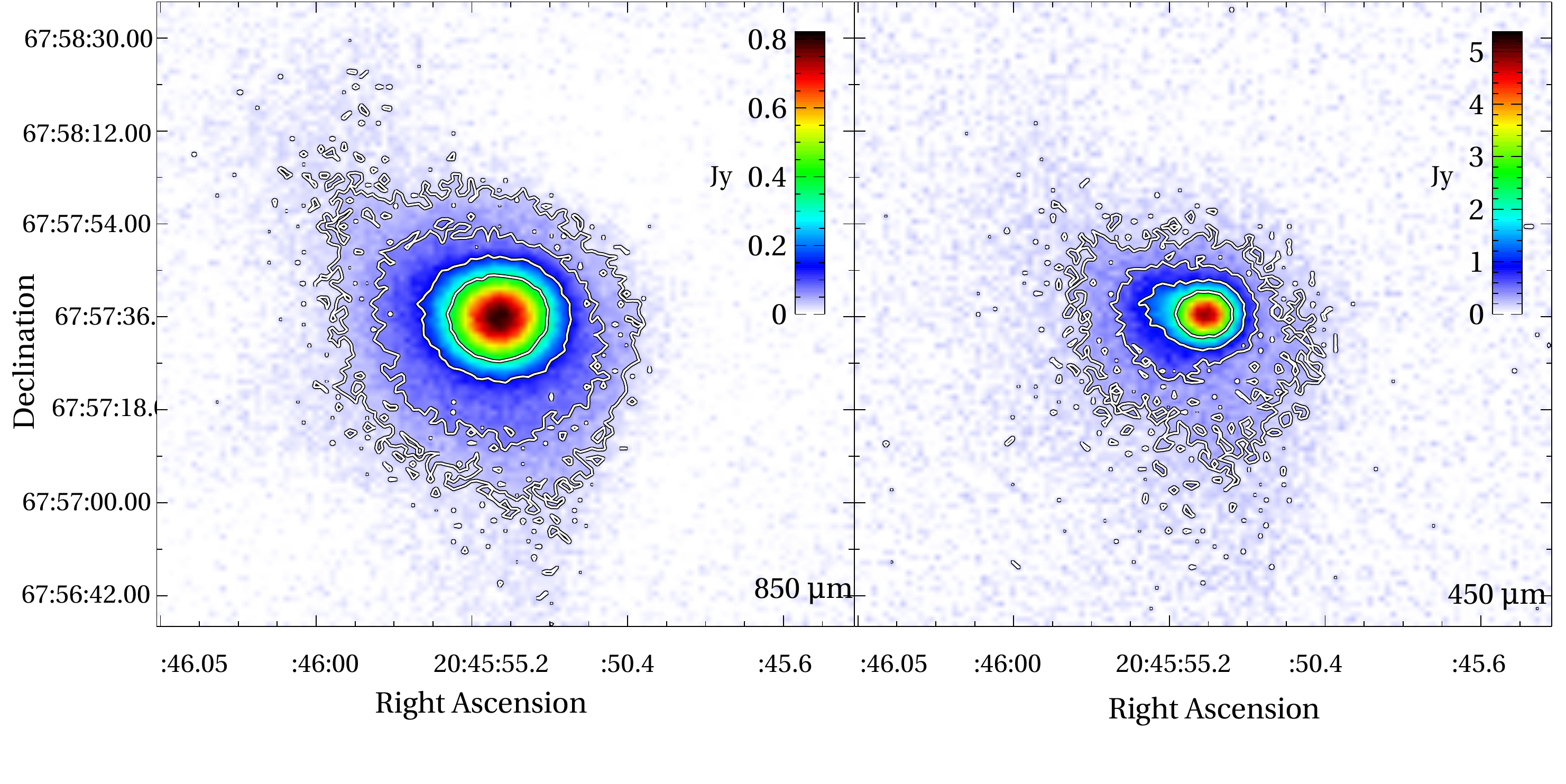}
   \caption[pvcep]
   { \label{fig:pvcep}
850\mum\ (left) and 450\mum\ (right) image of PV$\,$Cep. The colour scale is linear as shown in the colour bar on the right hand side. In each map, the five contours scale logarithmically from the peak signal as shown in Table~\ref{tab:newcals} down to the 3-$\sigma$ detection level (0.03/0.22$\,$Jy at 850/450\mum\ respectively).}
   \end{figure}

PV$\,$Cep is an irregular variable star, with strong extended emission at both SCUBA-2 wavelengths. This is reflected in the increased integrated flux in comparison with its peak value. It is, for these reasons, unlikely to be used as a standard calibrator. The peak fluxes at both wavelengths are significantly lower than the background-corrected flux densities reported by \cite{sandell2011} and this is quite likely a result of the variability of the source. \cite{sandell2011} also postulate that their background fitting at 450\mum\ might produce elevated 450\mum\ results. Figure~\ref{fig:pvcep} shows the well-resolved image of the circumstellar disk, and surrounding extended emission. PV$\,$Cep will not be a regularly observed calibrator as CRL$\,$2688 offers a better alternative in this part of the sky.\\

\subsection{MWC$\,$349}
\begin{figure}
   \includegraphics[height=4.2cm]{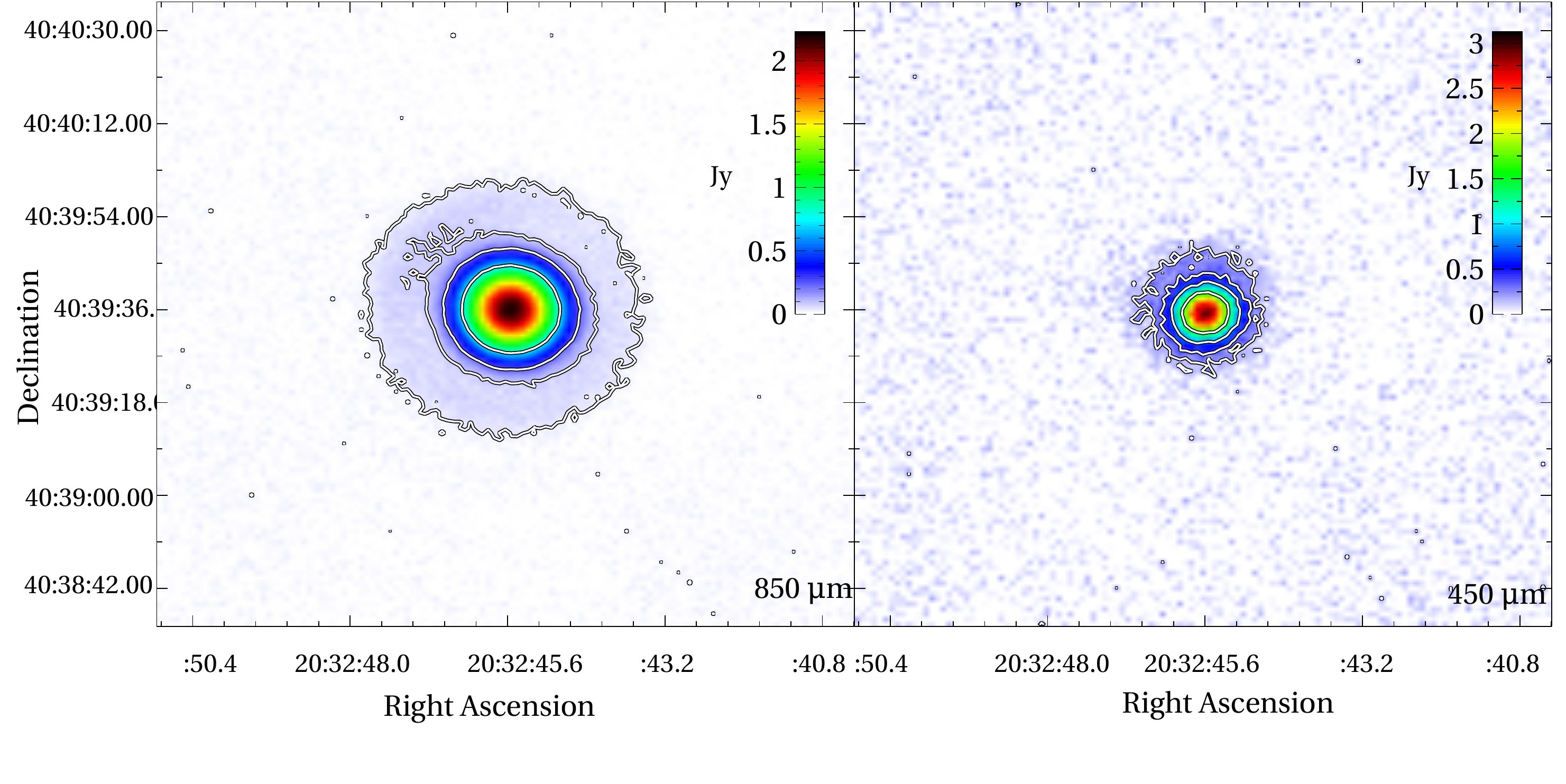}
   \caption[mwc349]
   { \label{fig:mwc349}
850\mum\ (left) and 450\mum\ (right) image of MWC$\,$349. The colour scale is linear as shown in the colour bar on the right hand side. In each map, the five contours scale linearly from the peak signal as shown in Table~\ref{tab:newcals} down to the 3-$\sigma$ detection level (0.03/0.18$\,$Jy at 850/450\mum\ respectively).}
   \end{figure}
MWC$\,$349, while quite close to CRL$\,$2688, has the advantage of being a point source object. Rejected as being too faint for calibration with SCUBA, it is satisfactorily detected at both wavelengths with SCUBA-2 as shown in Figure~\ref{fig:mwc349} and unresolved, with measured FWHM of 13.2 and 7.9 arc sec at 850/450\mum\ respectively, agreeing well with the main-beam Gaussian fit in Table~\ref{tab:gauss}. At 850\mum, we measure a flux of 2.19 $\pm$ 0.08 Jy, lower by a statistically significant amount to the 2.6 $\pm$ 0.07 Jy seen by \cite{sandell2011}. It is an emission-line star with strong free-free emission. We also detect a lower 450\mum\ flux of  3.2 $\pm$ 0.25$\,$Jy, in comparison to the 5.0 $\pm$ 1.1$\,$Jy SCUBA observation reported by \cite{sandell2011}. \cite{sandell2011} suggest that the background-subtraction method adopted with the SCUBA data produced questionably high 450\mum\ fluxes in some cases, though this does not explain the flux discrepancy at 850\mum. As these data were taken in a short period and with a long interval since the SCUBA measurements, source variability cannot be ruled out. CRL$\,$2688 remains a preferred calibrator to this source, which will not be regularly observed.\\

\subsection{HD$\,$169142}
\begin{figure}
   \includegraphics[height=4.2cm]{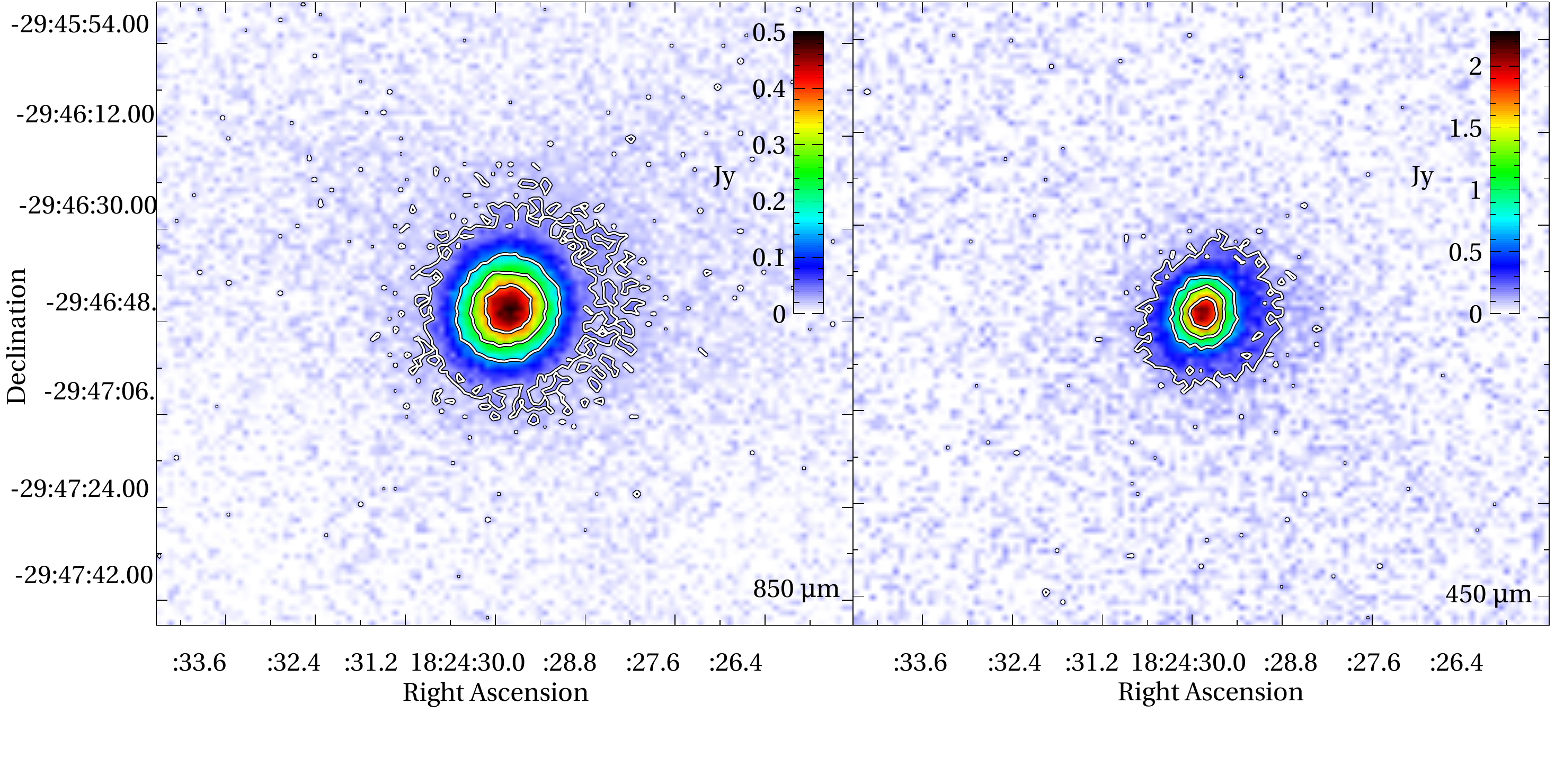}
   \caption[hd169142]
   { \label{fig:hd169142}
850\mum\ (left) and 450\mum\ (right) image of HD$\,$169142. The colour scale is linear as shown in the colour bar on the right hand side. In each map, the five contours scale linearly from the peak signal as shown in Table~\ref{tab:newcals} down to the 3-$\sigma$ detection level (0.03/0.17$\,$Jy at 850/450\mum\ respectively).}
   \end{figure}

HD$\,$169142 is a well-studied Herbig AeBe Star, and was selected as a study source for SCUBA-2 calibration to probe the sensitivity limits of the instrument, particularly at 450\mum. \cite{sandell2011} find that the disk may be resolved, and report a flux density of 0.57$\,$Jy at 850\mum\ and 3.34$\,$Jy at 450\mum. We measure an asymmetric disk with a fitted FWHM of 15.5$\times$13.4 arcsec at 850\mum\ which agrees with this finding. This is the faintest source observed with SCUBA-2 in this study and the 850\mum\ flux of 0.58$\,$Jy and 3.41$\,$Jy agrees with the value measured by \cite{sandell2011} within the uncertainty. The co-added maps of the source are shown in Figure~\ref{fig:hd169142}. \\

\subsection{V883$\,$Ori}
\begin{figure}
   \includegraphics[height=4.2cm]{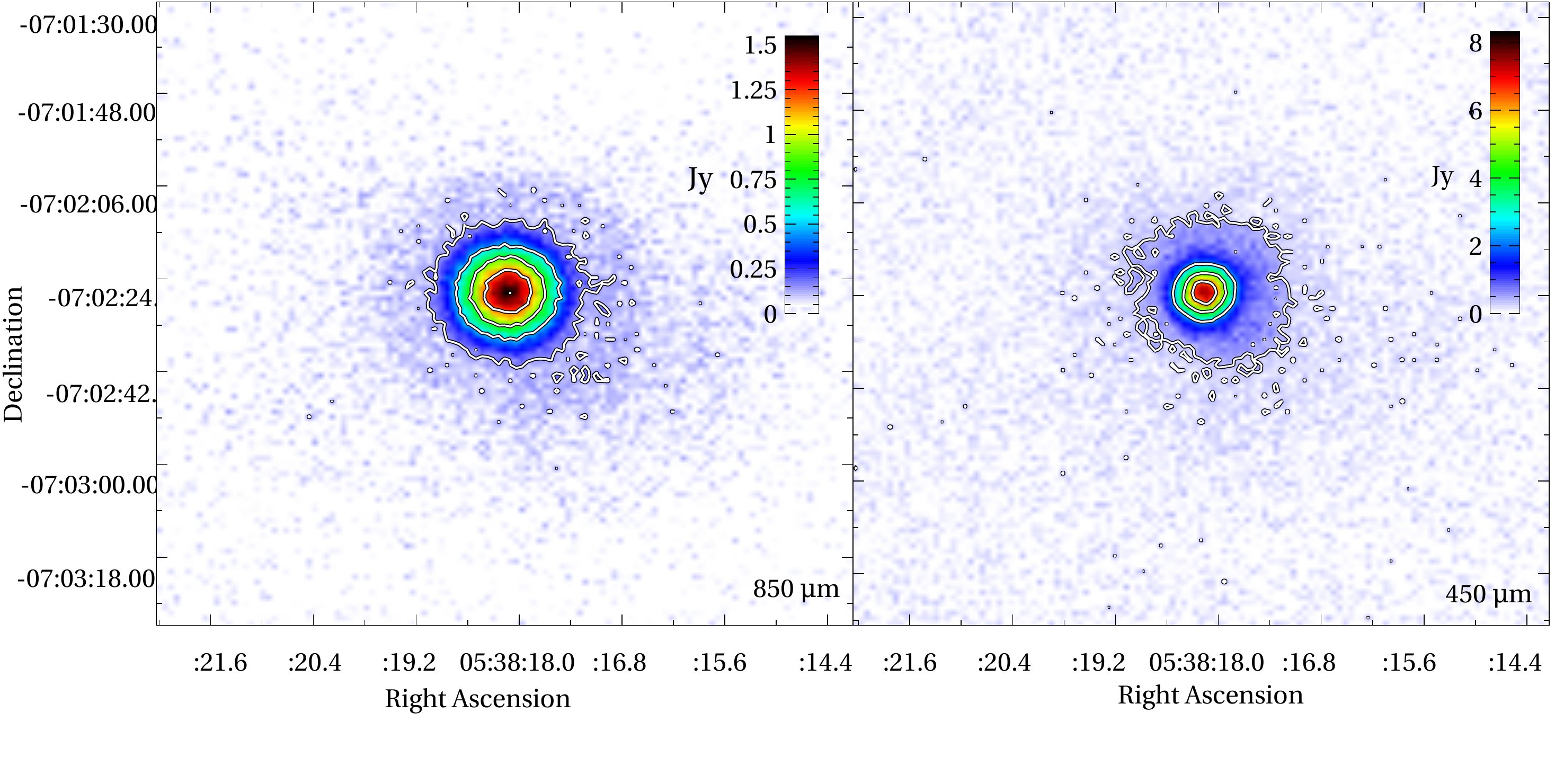}
   \caption[v883ori]
   { \label{fig:v883ori}
850\mum\ (left) and 450\mum\ (right) image of V883$\,$Ori. The colour scale is linear as shown in the colour bar on the right hand side. In each map, the five contours scale linearly from the peak signal as shown in Table~\ref{tab:newcals} down to the 3-$\sigma$ detection level (0.14/0.37$\,$Jy at 850/450\mum\ respectively).}
   \end{figure}

V883 Ori is FU Orionis star, which has a similar right ascension to a preferred calibrator, CRL$\,$618 but has the advantage of being in the south. \cite{sandell2001} observed this with SCUBA and found it to be unresolved, with little evident extended emission (though they suggest a possible emission region to the eastern edge of the 850\mum\ map). They measured a background-corrected flux density of 1.41 $\pm$ 0.02 Jy at 850\mum\,  and 9.45 $\pm$ 0.17 Jy at 450\mum. Additional observations with SCUBA measured an average peak flux at 450\mum\ flux of 7.28 $\pm$ 0.07 Jy \citep{barnard2}. We measure an 850\mum\ integrated flux of 2.0 $\pm$ 0.07 Jy and a 450\mum\ integrated flux of 10.4 $\pm$ 1 Jy. \cite{siringo} reports an integrated flux from LABOCA at 870\mum\ of 1.93 Jy, which agrees within our error range. The SCUBA-2 and LABOCA measurements both disagree with the lower 850\mum\ integrated flux measured by \cite{sandell2003} by a statistically significant margin.\\

\subsection{HL$\,$Tau}
\begin{figure}
   \includegraphics[height=4.2cm]{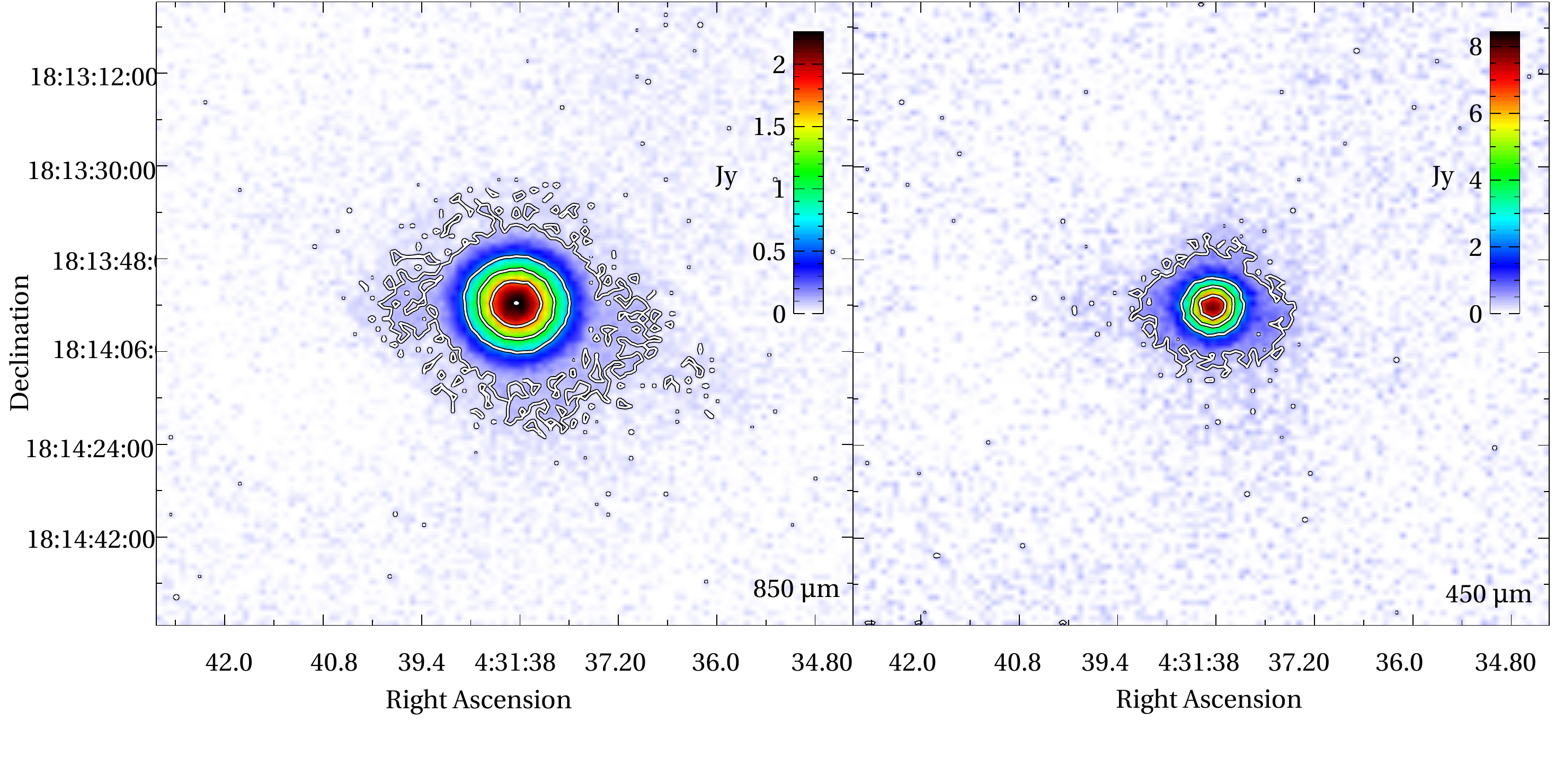}
   \caption[hltau]
   { \label{fig:hltau}
850\mum\ (left) and 450\mum\ (right) image of HL$\,$Tau. The colour scale is linear as shown in the colour bar on the right hand side. In each map, the five contours scale linearly from the peak signal as shown in Table~\ref{tab:newcals} down to the 3-$\sigma$ detection level (0.08/0.44$\,$Jy at 850/450\mum\ respectively).}
   \end{figure}

HL Tau is a young T Tauri star, and was used as a secondary calibrator for SCUBA \citep{sandell2003,jenness2002}. It is well studied in the submillimetre, particularly as imaging suggests a low-mass companion object in the parent disc material \citep{greaves2008}. The SCUBA-2 maps here do not detect a significant amount of this disc with a measured FWHM at 850\mum\ of 13.8 arc sec and 8.3 arc sec at 450\mum. The integrated fluxes agree well at both wavelengths with the measurements of \cite{jenness2002}. Only a small sample of data was taken of this source in the period available. Further observations will be conducted to explore the properties of the extended emission with better S/N. \\

\subsection{BVP$\,$1}
\begin{figure}
   \includegraphics[height=4.2cm]{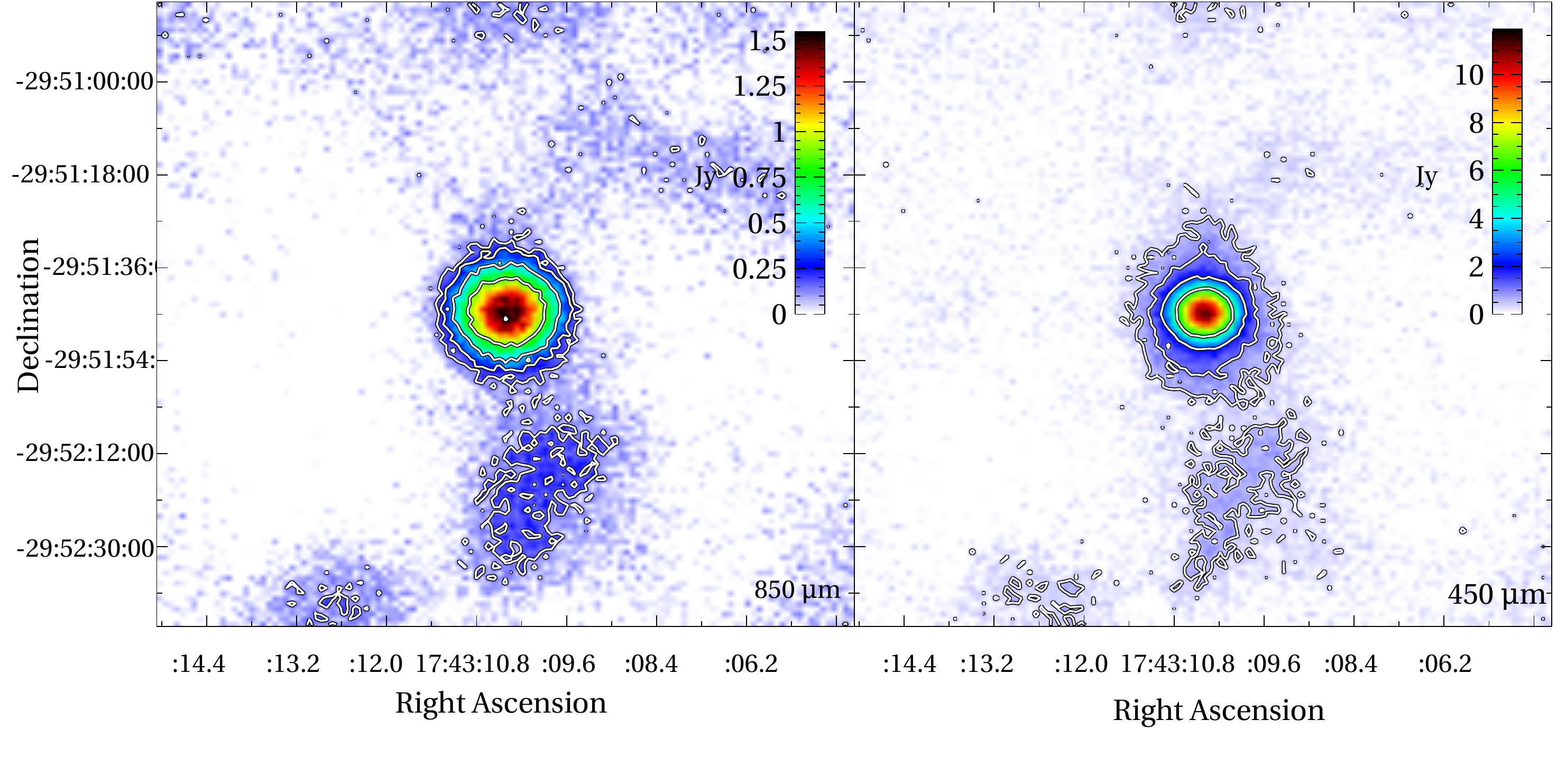}
   \caption[bvp1]
   { \label{fig:bvp1}
850\mum\ (left) and 450\mum\ (right) image of galactic source BVP$\,$1. The colour scale is linear as shown in the colour bar on the right hand side. In each map, the five contours scale logarithmically from the peak signal as shown in Table~\ref{tab:newcals} down to the 3-$\sigma$ detection level (0.144/0.46$\,$Jy at 850/450\mum\ respectively). The correlated background structures are likely a result of the proximity of the source to the galactic centre.}
   \end{figure}
Figure~\ref{fig:bvp1} shows the SCUBA-2 images of BVP$\,$1, a bright, compact submillimetre source located near the galactic centre. SCUBA observations of the source are to date the only published results at these wavelengths, and \cite{barnard} measured a peak flux at 1.069/11.318$\,$Jy at 850/450\mum\ respectively. It is a good potential secondary calibrator for SCUBA-2 as there is a paucity of good calibration candidates in this region of the sky. With only a handful of observations to date, SCUBA-2 measures peak fluxes of 1.37/11.9$\,$Jy at 850/450\mum\ respectively. This is higher at both wavelengths than \cite{barnard}, with a 4-$\sigma$ discrepancy at 850\mum. The integrated fluxes are 1.55/16.9$\,$Jy at 850/450\mum. Fits of the source profile show that it is compact at 850\mum, with a central FWHM of 14.3 arcsec. It shows extended emission at 450\mum, as indicated by the elevated ratio of integrated to peak flux, though its central core is extremely bright and fairly compact with a measured FWHM of 8.8$\,$arcsec. The SCUBA observations were taken in jiggle-map mode, and in the crowded region near the galactic centre, with large amounts of extended emission, it is possible that chopped emission resulted in the lower flux recorded by \cite{barnard}. Observations and investigation of this object continue and will hopefully shed more light its properties.\\

\section{Conclusions}

SCUBA-2 is fully commissioned and currently running very successful science operations on the JCMT. Atmospheric calibration at both wavelengths has been optimised using new line-of-sight opacity algorithms and the derived atmospheric extinction relations are shown here. Using a  large sample of primary and secondary calibrator observations, the flux conversion factors (FCF) for the 850\mum\ and 450\mum\ arrays are deduced and an investigation of the instrument beam-shape and photometry methods is given.  The relative flux calibration is better than 10 per cent at both wavelengths. The sample-size of the calibration observations and accurate FCFs have allowed the determination of the 850\mum\ and 450\mum\ fluxes of several well-known submillimetre sources and these results are presented.\\

\section*{Acknowledgments}
The James Clerk Maxwell Telescope is operated by the Joint Astronomy Centre on behalf of the Science and Technology Facilities Council of the United Kingdom, the Netherlands Organisation for Scientific Research, and the National Research Council of Canada. Additional funds for the construction of SCUBA-2 were provided by the Canada Foundation for Innovation. The authors would like to thank Doug Johnstone for his comments and input to this paper.\\

\bibliography{scuba2}  

\begin{thebibliography}{}

\bibitem[\protect\citeauthoryear{{Aguirre} et~al.,}{{Aguirre}
  et~al.}{2011}]{aguirre2011}
{Aguirre} J.~E.,  et~al., 2011, ApJS, 192, 4

\bibitem[\protect\citeauthoryear{{Archibald} et~al.,}{{Archibald}
  et~al.}{2002}]{archibald}
{Archibald} E.~N.,  et~al., 2002, MNRAS, 336, 1

\bibitem[\protect\citeauthoryear{{Barnard}}{{Barnard}}{2005}]{barnard2}
{Barnard} V.~E.,  2005, Technical Report {SCD/SN/011}, {A summary of the search
  for new Secondary Submillimetre Calibrators}.
JCMT

\bibitem[\protect\citeauthoryear{{Barnard}, {Vielva}, {Pierce-Price}, {Blain},
  {Barreiro}, {Richer} \& {Qualtrough}}{{Barnard} et~al.}{2004}]{barnard}
{Barnard} V.~E.,  {Vielva} P.,  {Pierce-Price} D.~P.~I.,  {Blain} A.~W.,
  {Barreiro} R.~B.,  {Richer} J.~S.,    {Qualtrough} C.,  2004, MNRAS, 352, 961

\bibitem[\protect\citeauthoryear{{Bintley} et~al.,}{{Bintley}
  et~al.}{2012}]{bintley2012}
{Bintley} D.,  et~al., 2012, in {Holland} W.~S.,  {Zmuindzinas} J.,  eds,  SPIE
  Conf. Series Vol. 8452, Millimeter, Submillimeter, and Far-Infrared Detectors
  and Instrumentation for Astronomy VI. p.~19

\bibitem[\protect\citeauthoryear{{Bintley} et~al.,}{{Bintley}
  et~al.}{2010}]{bintley2010}
{Bintley} D.,  et~al., 2010, in {Holland} W.~S.,  {Zmuindzinas} J.,  eds,  SPIE
  Conf. Series Vol. 7741, Millimeter, Submillimeter, and Far-Infrared Detectors
  and Instrumentation for Astronomy V. p.~1

\bibitem[\protect\citeauthoryear{{Buckle} et~al.,}{{Buckle}
  et~al.}{2009}]{buckle}
{Buckle} J.~V.,  et~al., 2009, MNRAS, 399, 1026

\bibitem[\protect\citeauthoryear{{Chapin}, {Berry}, {Gibb}, {Jenness}, {Scott}
  \& {Economou}}{{Chapin} et~al.}{2013}]{chapin}
{Chapin} E.~L.,  {Berry} D.~S.,  {Gibb} A.~G.,  {Jenness} T.,  {Scott} D.,
  {Economou} F.,  2013, MNRAS, {In press}

\bibitem[\protect\citeauthoryear{{Currie}, {Draper}, {Berry}, {Jenness},
  {Cavanagh} \& {Economou}}{{Currie} et~al.}{2008}]{currie}
{Currie} M.~J.,  {Draper} P.~W.,  {Berry} D.~S.,  {Jenness} T.,  {Cavanagh} B.,
     {Economou} F.,  2008, in {Argyle} R.~W.,  {Bunclark} P.~S.,   {Lewis}
  J.~R.,  eds,  Astron. Soc. of the Pac. Conf. Series Vol. 394, Astronomical
  Data Analysis Software and Systems XVII. p.~650

\bibitem[\protect\citeauthoryear{{Dempsey} \& {Friberg}}{{Dempsey} \&
  {Friberg}}{2008}]{dempsey2008}
{Dempsey} J.~T.,  {Friberg} P.,  2008, in {Stepp} L.~M.,  {Gilmozzi} R.,  eds,
  SPIE Conference Series Vol. 7012, Ground-based and Airborne Telescopes II. pp
  70123Z--20123Z--11

\bibitem[\protect\citeauthoryear{{Dempsey} et~al.,}{{Dempsey}
  et~al.}{2012}]{dempsey2012}
{Dempsey} J.~T.,  et~al., 2012, in {Holland} W.~S.,  {Zmuindzinas} J.,  eds,
  SPIE Conf. Series Vol. 8452, Millimeter, Submillimeter, and Far-Infrared
  Detectors and Instrumentation for Astronomy VI. pp 1--18

\bibitem[\protect\citeauthoryear{{Dunne}, {Eales}, {Edmunds}, {Ivison},
  {Alexander} \& {Clements}}{{Dunne} et~al.}{2000}]{dunne}
{Dunne} L.,  {Eales} S.,  {Edmunds} M.,  {Ivison} R.,  {Alexander} P.,
  {Clements} D.~L.,  2000, MNRAS, 315, 115

\bibitem[\protect\citeauthoryear{{Greaves}, {Richards}, {Rice} \&
  {Muxlow}}{{Greaves} et~al.}{2008}]{greaves2008}
{Greaves} J.~S.,  {Richards} A.~M.~S.,  {Rice} W.~K.~M.,    {Muxlow} T.~W.~B.,
  2008, MNRAS, 391, L74

\bibitem[\protect\citeauthoryear{{Holland} et~al.,}{{Holland}
  et~al.}{2013}]{holland2012}
{Holland} W.~S.,  et~al., 2013, MNRAS, {In press}

\bibitem[\protect\citeauthoryear{{Holland} et~al.,}{{Holland}
  et~al.}{1999}]{holland1999}
{Holland} W.~S.,  et~al., 1999, MNRAS, 303, 659

\bibitem[\protect\citeauthoryear{{Jenness}, {Berry}, {Chapin}, {Economou},
  {Gibb} \& {Scott}}{{Jenness} et~al.}{2011}]{jenness2011}
{Jenness} T.,  {Berry} D.,  {Chapin} E.,  {Economou} F.,  {Gibb} A.,    {Scott}
  D.,  2011, in {Evans} I.~N.,  {Accomazzi} A.,  {Mink} D.~J.,   {Rots} A.~H.,
  eds,  Astronomical Society of the Pacific Conference Series Vol. 442,
  Astronomical Data Analysis Software and Systems XX. p.~281

\bibitem[\protect\citeauthoryear{{Jenness}, {Robson} \& {Stevens}}{{Jenness}
  et~al.}{2010}]{jenness2010}
{Jenness} T.,  {Robson} E.~I.,    {Stevens} J.~A.,  2010, MNRAS, 401, 1240

\bibitem[\protect\citeauthoryear{{Jenness}, {Stevens}, {Archibald}, {Economou},
  {Jessop} \& {Robson}}{{Jenness} et~al.}{2002}]{jenness2002}
{Jenness} T.,  {Stevens} J.~A.,  {Archibald} E.~N.,  {Economou} F.,  {Jessop}
  N.~E.,    {Robson} E.~I.,  2002, MNRAS, 336, 14

\bibitem[\protect\citeauthoryear{{Knapp}, {Sandell} \& {Robson}}{{Knapp}
  et~al.}{1993}]{knapp}
{Knapp} G.~R.,  {Sandell} G.,    {Robson} E.~I.,  1993, ApJS, 88, 173

\bibitem[\protect\citeauthoryear{{Kov{\'a}cs}, {Chapman}, {Dowell}, {Blain},
  {Ivison}, {Smail} \& {Phillips}}{{Kov{\'a}cs} et~al.}{2006}]{kovacs}
{Kov{\'a}cs} A.,  {Chapman} S.~C.,  {Dowell} C.~D.,  {Blain} A.~W.,  {Ivison}
  R.~J.,  {Smail} I.,    {Phillips} T.~G.,  2006, ApJ, 650, 592

\bibitem[\protect\citeauthoryear{{Lisenfeld}, {Isaak} \& {Hills}}{{Lisenfeld}
  et~al.}{2000}]{lisenfeld}
{Lisenfeld} U.,  {Isaak} K.~G.,    {Hills} R.,  2000, MNRAS, 312, 433

\bibitem[\protect\citeauthoryear{{Privett}, {Jenness}, {Matthews} \&
  {Barnard}}{{Privett} et~al.}{2005}]{privett}
{Privett} G.,  {Jenness} T.,  {Matthews} H.,    {Barnard} V.,  2005, {Fluxes --
  JCMT Position and Flux Densitivy Calibration}.
Starlink User Note 213, Starlink Project, CLRC

\bibitem[\protect\citeauthoryear{{Radford}}{{Radford}}{2011}]{radford}
{Radford} S.~J.~E.,  2011, in {Cure} M.,  {Otarola} A.,  {Marin} J.,
  {Sarazin} M.,  eds,  Revista Mexicana de Astronomia y Astrofisica Conference
  Series Vol. 41, Astronomical Site Testing Data in Chile. p.~87

\bibitem[\protect\citeauthoryear{{Sakamoto} et~al.,}{{Sakamoto}
  et~al.}{2008}]{sakamoto}
{Sakamoto} K.,  et~al., 2008, ApJ, 684, 957

\bibitem[\protect\citeauthoryear{{Sandell}}{{Sandell}}{2003}]{sandell2003}
{Sandell} G.,  2003, in {Metcalfe} L.,  {Salama} A.,  {Peschke} S.~B.,
  {Kessler} M.~F.,  eds,  ESA Special Publication Vol. 481, The Calibration
  Legacy of the ISO Mission. p.~439

\bibitem[\protect\citeauthoryear{{Sandell} \& {Weintraub}}{{Sandell} \&
  {Weintraub}}{2001}]{sandell2001}
{Sandell} G.,  {Weintraub} D.~A.,  2001, ApJS, 134, 115

\bibitem[\protect\citeauthoryear{{Sandell}, {Weintraub} \&
  {Hamidouche}}{{Sandell} et~al.}{2011}]{sandell2011}
{Sandell} G.,  {Weintraub} D.~A.,    {Hamidouche} M.,  2011, ApJ, 727, 26

\bibitem[\protect\citeauthoryear{{Sayers} et~al.,}{{Sayers}
  et~al.}{2010}]{sayers}
{Sayers} J.,  et~al., 2010, ApJ, 708, 1674

\bibitem[\protect\citeauthoryear{{Seiffert}, {Borys}, {Scott} \&
  {Halpern}}{{Seiffert} et~al.}{2007}]{seiffert}
{Seiffert} M.,  {Borys} C.,  {Scott} D.,    {Halpern} M.,  2007, MNRAS, 374,
  409

\bibitem[\protect\citeauthoryear{{Siringo} et~al.,}{{Siringo}
  et~al.}{2009}]{siringo}
{Siringo} G.,  et~al., 2009, A\&A, 497, 945

\bibitem[\protect\citeauthoryear{{Stevens} \& {Robson}}{{Stevens} \&
  {Robson}}{1994}]{stevens}
{Stevens} J.~A.,  {Robson} E.~I.,  1994, MNRAS, 270, L75

\bibitem[\protect\citeauthoryear{{Truch} et~al.,}{{Truch} et~al.}{2008}]{truch}
{Truch} M.~D.~P.,  et~al., 2008, ApJ, 681, 415

\bibitem[\protect\citeauthoryear{{Wiedner}, {Hills}, {Carlstrom} \&
  {Lay}}{{Wiedner} et~al.}{2001}]{wiedner}
{Wiedner} M.~C.,  {Hills} R.~E.,  {Carlstrom} J.~E.,    {Lay} O.~P.,  2001,
  ApJ, 553, 1036

\bibitem[\protect\citeauthoryear{{Wright}}{{Wright}}{1976}]{wright1976}
{Wright} E.~L.,  1976, ApJ, 210, 250

\end{thebibliography}
\bibliographystyle{mn2e}

\label{lastpage}

\end{document}